\documentclass{article}
\pdfoutput=1
\usepackage{graphicx}
\usepackage{subfig}
\usepackage{amsmath,amsfonts}
\usepackage{bm}
\usepackage{xcolor}
\usepackage{tensor}
\usepackage{url}
\newcommand{\argmin}{\operatornamewithlimits{arg\ min}}
\newcommand{\tauij}{\tau_{ij}}
\newcommand{\aij}{a_{ij}}

\newcommand{\bij}{b_{ij}}

\newcommand{\bdelta}{b^\Delta_{ij}}
\newcommand{\bpk}{b^R_{ij}}
\newcommand{\Sij}{S_{ij}}
\newcommand{\Oij}{\Omega_{ij}}
\newcommand{\bmR}{\bm{R}}
\newcommand{\bmbdelta}{\bm{b^\Delta}}

\newcommand{\norm}[1]{\left\lVert #1 \right\rVert}
\newcommand{\komegasst}{$k$-$\omega$ SST}
\newcommand{\B}{\bm{\mathcal{B}}}
\newcommand{\C}[1]{\bm{\mathcal{C}}_{#1}}
\newcommand{\Cd}[1]{\bm{C}_{#1}}
\newcommand{\bmtheta}{\bm{\Theta}}
\newcommand{\bmdelta}{\bm{\Delta}}

\newcommand{\method}{SpaRTA} 
\newcommand{\frozen}{$k$-corrective-frozen-RANS}
\newcommand{\T}[1]{T^{(#1)}_{ij}}
\newcommand{\cbfs}[1]{$\text{CBFS}_{13700}$#1}
\newcommand{\ph}[1]{$\text{PH}_{10595}$#1}
\newcommand{\phexp}[1]{$\text{PH}_{37000}$#1}
\newcommand{\cd}[1]{$\text{CD}_{12600}$#1}

\newcommand{\m}[2]{M^{#1}_{#2}}

\begin{document}

\title{Discovery of Algebraic Reynolds-Stress Models Using Sparse Symbolic Regression.}

\author{Martin Schmelzer  \and
        Richard P. Dwight \and
        Paola Cinnella}

\date{\vspace{1cm}
\raggedright{ This article is published (open-access). Please, use the original article instead of this one. For citation:
\\
\vspace{0.5cm}
\bfseries{Schmelzer, M., Dwight, R.P. \& Cinnella, P. 
\\
Discovery of Algebraic Reynolds-Stress Models Using Sparse Symbolic Regression. 
\\
Flow Turbulence Combustion 104, 579–603 (2020). 
\\
\url{https://doi.org/10.1007/s10494-019-00089-x}
\\}
}}

\maketitle

\begin{abstract}
A novel deterministic symbolic regression method \method{} is introduced to infer algebraic stress models for the closure of RANS equations directly from high-fidelity LES or DNS data. The models are written as tensor polynomials and are built from a library of candidate functions. The machine-learning method is based on elastic net regularisation which promotes sparsity of the inferred models. By being data-driven the method relaxes assumptions commonly made in the process of model development. Model-discovery and cross-validation is performed for three cases of separating flows, i.e.~periodic hills ($Re$=10595), converging-diverging channel ($Re$=12600) and curved backward-facing step ($Re$=13700). The predictions of the discovered models are significantly improved over the \komegasst{} also for a \emph{true} prediction of the flow over periodic hills at $Re$=37000. This study shows a systematic assessment of \method{} for rapid machine-learning of robust corrections for standard RANS turbulence models.

\end{abstract}

\section{Introduction}
 The capability of Computational Fluid Dynamics (CFD) to deliver reliable prediction is limited by the unsolved closure problem of turbulence modelling. The workhorse for turbulence modelling in industry are the Reynolds-Averaged Navier-Stokes (RANS) equations using linear eddy viscosity models (LEVM) \cite{Slotnick2014}. The lower computational costs compared to high-fidelity approaches, e.g. Large-Eddy (LES) or Direct Numerical Simulations (DNS), come at the price of uncertainty especially for flows with separation, adverse pressure gradients or high streamline curvature. Data-driven methods for turbulence modelling based on supervised machine learning have been introduced to leverage RANS for improved predictions \cite{Xiao2019,Duraisamy2019,Durbin2018}. In \cite{Tracey2015a}, the source terms of the Spalart-Allmaras were learnt from data using a single hidden layer neural network, which served as a first feasibility study. In \cite{Parish2016}, a factor was introduced to correct the turbulent production in the $k$-equation of the $k$-$\omega$ model. This term was found via inverse modelling and served to train a Gaussian process. While this approach has been extended and applied to industrially relevant flows such as airfoils in \cite{Singh2016,Singh2017a} it still relies on the Boussinesq assumption. In \cite{Ling2016b}, a deep neural network was trained to predict $\aij$ given input only from a baseline linear eddy viscosity simulation and thus replacing the turbulence model instead of augmenting it. The network was designed to embed Galilean invariance of the predicted $\aij$. This concept of physics-informed machine learning was extended, e.g., in \cite{Wu2018} using random forest regression. Despite the success of the data-driven approaches a drawback is their black box nature, which hampers the understanding of the physics of the resulting models in order to derive new modelling ideas from it. 
\begin{figure}
\centering
  \includegraphics[width=0.75\textwidth]{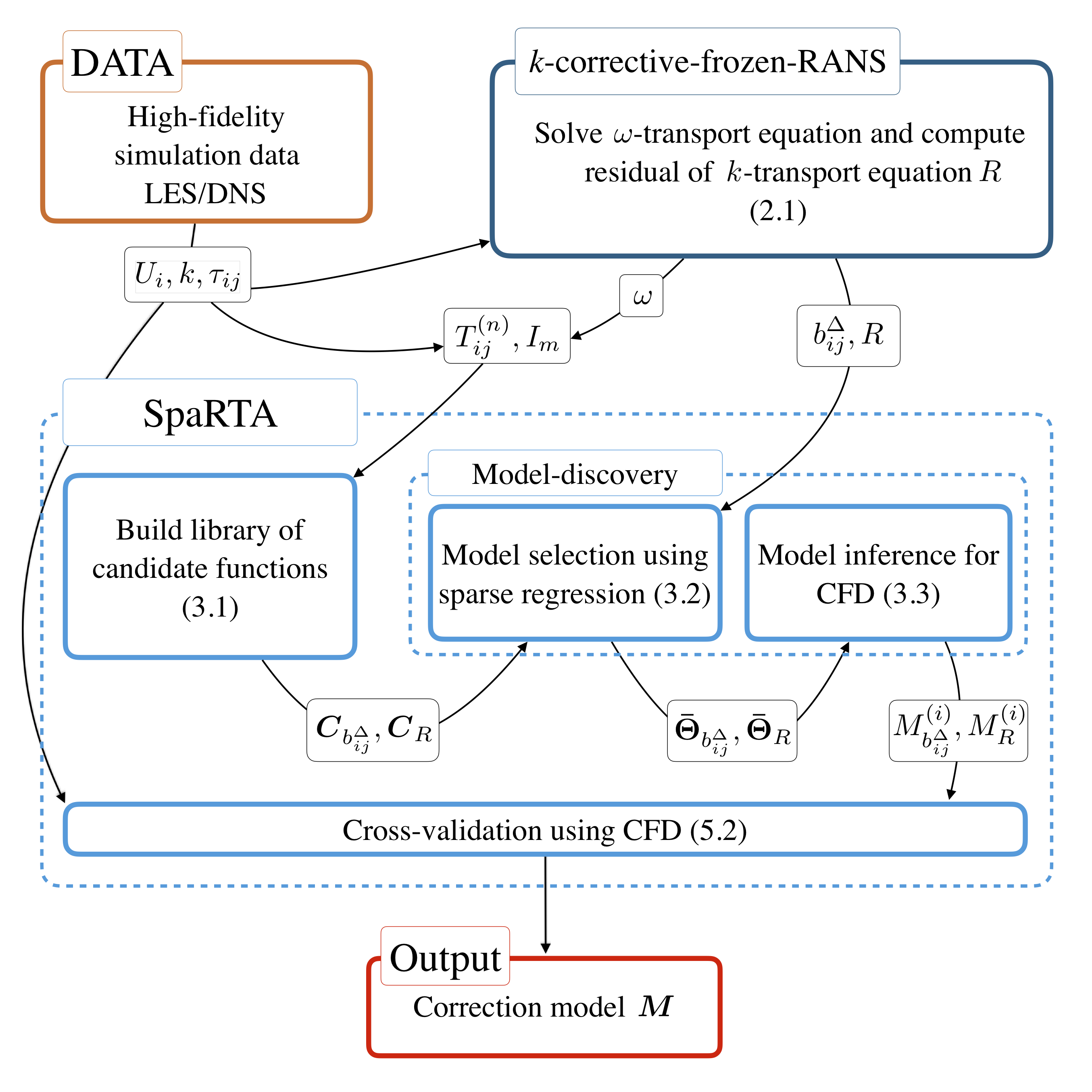}
\caption{Technical flow diagram of \method{} (\underline{Spa}rse \underline{R}egression of \underline{T}urbulent Stress \underline{A}nisotropy).}
\label{fig:diagram}       
\end{figure}%

Recently, a method has been introduced using genetic-programming (GEP) based symbolic regression to derive Explicit Algebraic Reynolds-stress Models (EARSM) directly from high-fidelity data \cite{Weatheritt2016,Weatheritt2017b}. EARSM, first introduced by \cite{Pope1975} and further developed by \cite{Gatski1993}, are nonlinear extensions of LEVM and are commonly derived by projecting Reynolds-stress models (RSM) onto a set of tensorial polynomials \cite{Leschziner2015,Pope2000}. These models are numerically more robust than RSM at similar computational costs as LEVM \cite{Wallin2000}, but do not show superior predictive capabilities for all kinds of flows \cite{Leschziner2015}. The data-driven GEP method retains the input quantities used to derive EARSM, but replaces the commonly used projection method to find the formal structure of the model by an evolutionary process, which makes it an open-box machine learning approach. The advantage of such a data-driven method is that instead of relying on assumptions made during the development of an EARSM, a model is inferred directly from data. While such a model might not provide an universal approach for all kinds of flows as commonly aimed for in physical modelling, it serves as a pragmatic tool to correct the flow at hand. For cases exhibiting similar flow physics, e.g.~separation, it has also been shown that the discovered models provide suitable corrections indicating the predictive potential of a data-driven approach. 

Due to the non-deterministic nature of GEP it discovers for each run another model with a different mathematical form, e.g.~other terms and/or other values for coefficients, with varying complexity. It is reported that the models using only a few nonlinear terms show a low training and prediction error as well as high numerical robustness for industrially relevant flow cases \cite{Weatheritt2017a,Akolekar2018}. Therefore, we instead introduce a new deterministic symbolic regression method \method{} (\underline{Spa}rse \underline{R}egression of \underline{T}urbulent Stress \underline{A}nisotropy), for which we constrain the search towards sparse algebraic models using sparsity-promoting regression techniques \cite{Brunton2016,Rudy2017}. \method{} combines functions from a predefined library of candidates without any random recombination. It consists of four steps: (i) building a library of candidate functions, (ii) model selection using sparse-regression techniques, (iii) inference of model coefficients and (iv) cross-validation of the resulting models, see Figure \ref{fig:diagram}. The first three steps are computationally very cheap also for high-dimensional problems and allow for rapid model discovery.

The present study provides several novel concepts for data-driven modelling, which are organised as follows. In Section 2 we define additive model-form error terms within the \komegasst{} LEVM model and use \frozen{}, which is an extension of the method introduced in \cite{Weatheritt2017b}, to compute the model-form error from high-fidelity data. The novelty in this work is that we identify not only a correction of the stress-strain relation, but also one for the turbulent transport equations and thereby achieve excellent agreement with mean-fields of high-fidelity data. We also validate that the model-form error is successfully captured by adding the two terms to the solver and performing a CFD simulation. The \frozen{} does not require any iterative optimisation procedure as compared to \cite{Parish2016} and is therefore very efficient, but also limited to full-field data. In Section 3 we introduce the steps of \method{}. The details of the test cases, the CFD setup and the sources of the high-fidelity data are given in Section 4. In Section 5 \method{} is applied to the test cases, the discovered models are presented and the best models are chosen using cross-validation. Finally, conclusions are drawn in Section 6.

\section{Model-form error of RANS equations}
\label{sec:model-form-error}
In the following, we augment the baseline model, i.e.~the linear eddy viscosity assumption and the turbulence transport equations of the \komegasst{}, with additive terms accounting for the error due to the model-form. We introduce \frozen{}, which is an extension of the method in \cite{Weatheritt2017b}, to extract these two types of error from high-fidelity data sources efficiently. Finally, we validate that the extracted terms reduce the error for given test cases.

\subsection{Identification of additive model-form error from data}
\label{subsec::additive_modelform_error}
The incompressible and constant-density RANS equations read
\begin{align}
	\partial_i U_i &= 0, \nonumber \\
	U_j \partial_j U_i &= \partial_j \left[ -\frac{1}{\rho} P + \nu \partial_j U_i - \tauij \right],
\end{align}
\noindent where $U_i$ is the mean velocity, $\rho$ is the constant density, $P$ is the mean pressure and $\nu$ is the kinematic viscosity. The Reynolds-stress $\tauij$ is the subject of modelling. This symmetric, second-order tensor field can be decomposed into an anisotropic $\aij = 2k\bij$ and isotropic part $\frac{2}{3} k \delta_{ij}$
\begin{align}
	\tauij &= 2k \left( \bij  + \frac{1}{3} \delta_{ij} \right), \label{eq::reynoldsstress}
\end{align}

\noindent in which the baseline model, $\bij^o = -\frac{\nu_t}{k} S_{ij}$, forms a linear relation between anisotropy and the mean-strain rate tensor $\Sij$ via the scalar eddy viscosity $\nu_t$. Commonly, $\nu_t$ is computed using a transport model such as \komegasst{} \cite{Leschziner2015}, in which $k$ is the turbulent kinetic energy and $\omega$ the specific dissipation rate. 

In order to extract the model-form error in these models from high-fidelity data sources, we compute the residuals of the baseline turbulence model given the data. The residual for the constitutive relation is equivalent to an additive term $\bdelta$ leading to an augmented constitutive relation
\begin{align}
	\bij   &= -\frac{\nu_t}{k} S_{ij} + \bdelta.\label{eq::nonlinearconstrel}
\end{align}

\begin{figure}
	\centering
	\subfloat[\ph{}\label{fig:verify:ph}]{%
       \includegraphics[width=0.85\textwidth]{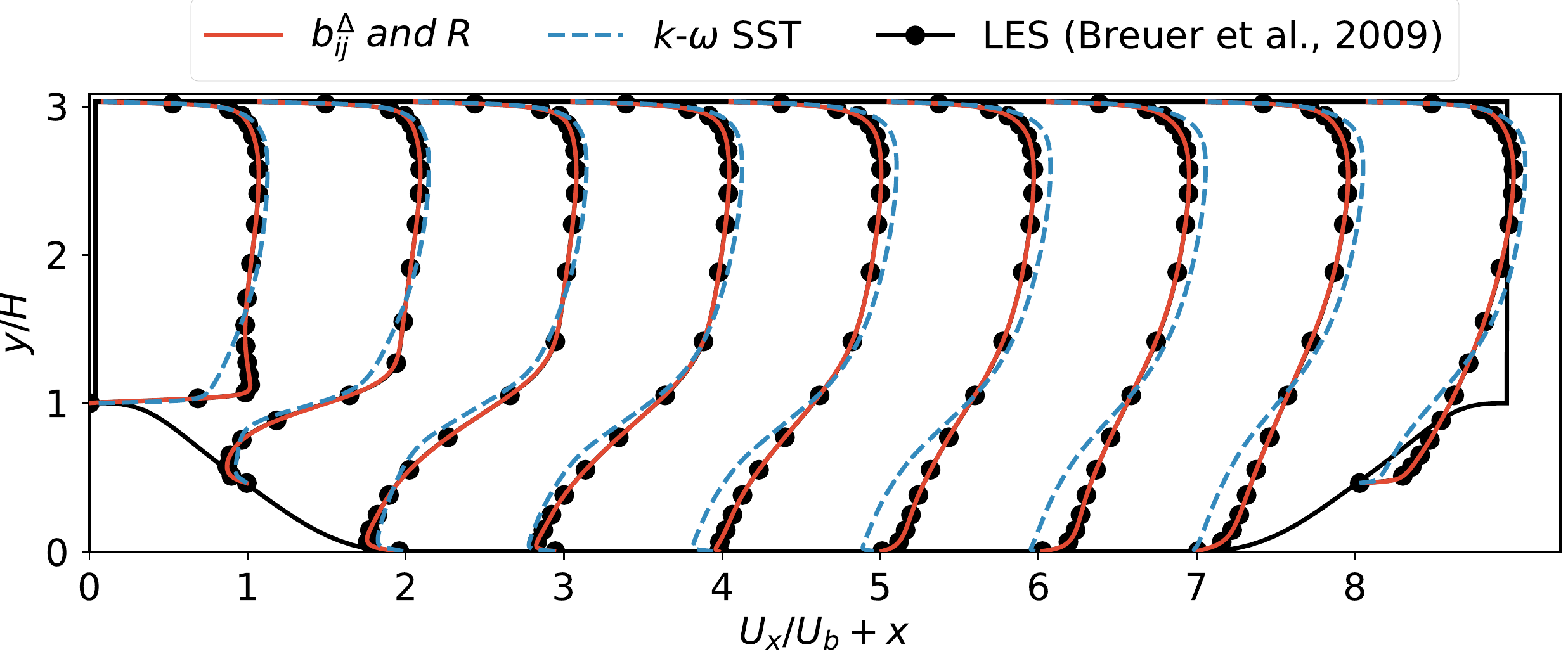}
    }
    \hfill
    \subfloat[\cd{}\label{fig:verify:cd}]{%
		\includegraphics[width=0.85\textwidth]{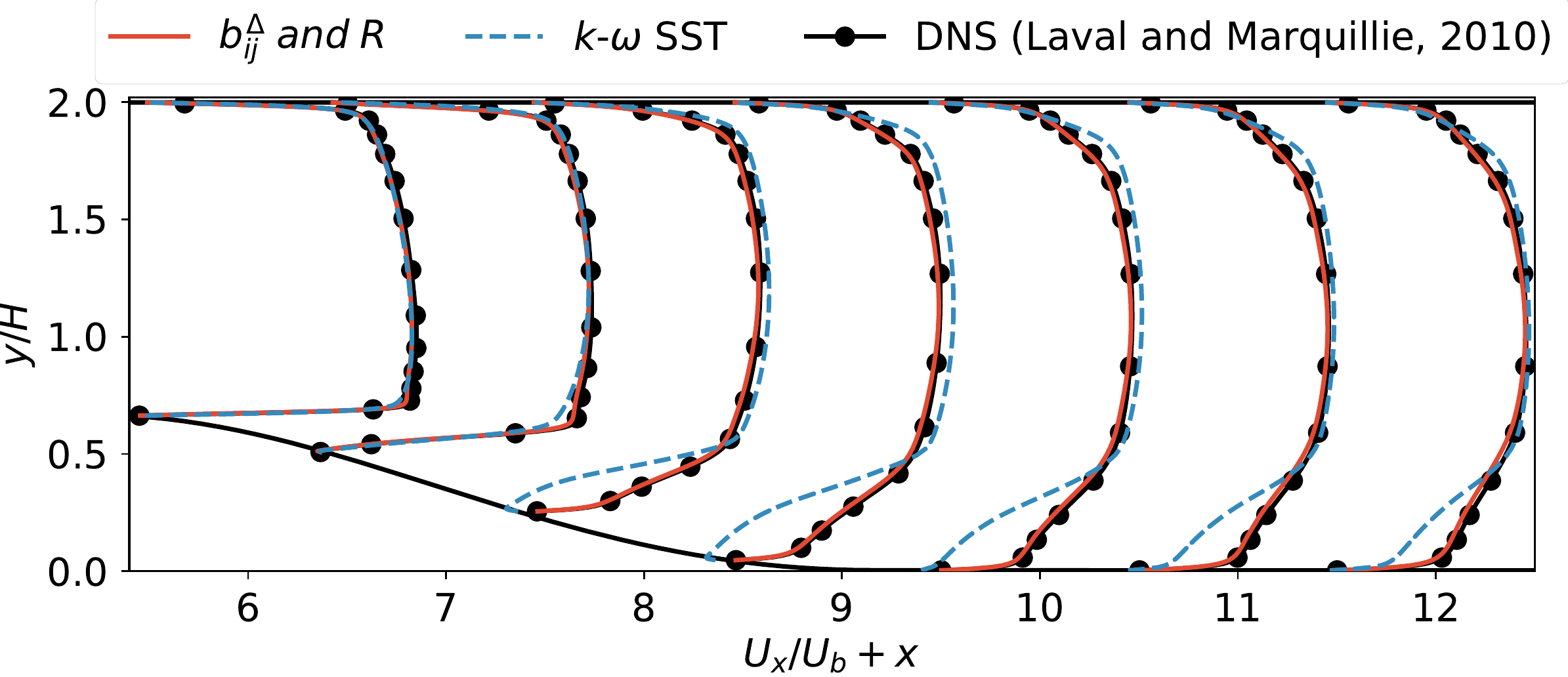}
    }
	\hfill
    \subfloat[\cbfs{}\label{fig:verify:cbfs}]{%
		\includegraphics[width=0.85\textwidth]{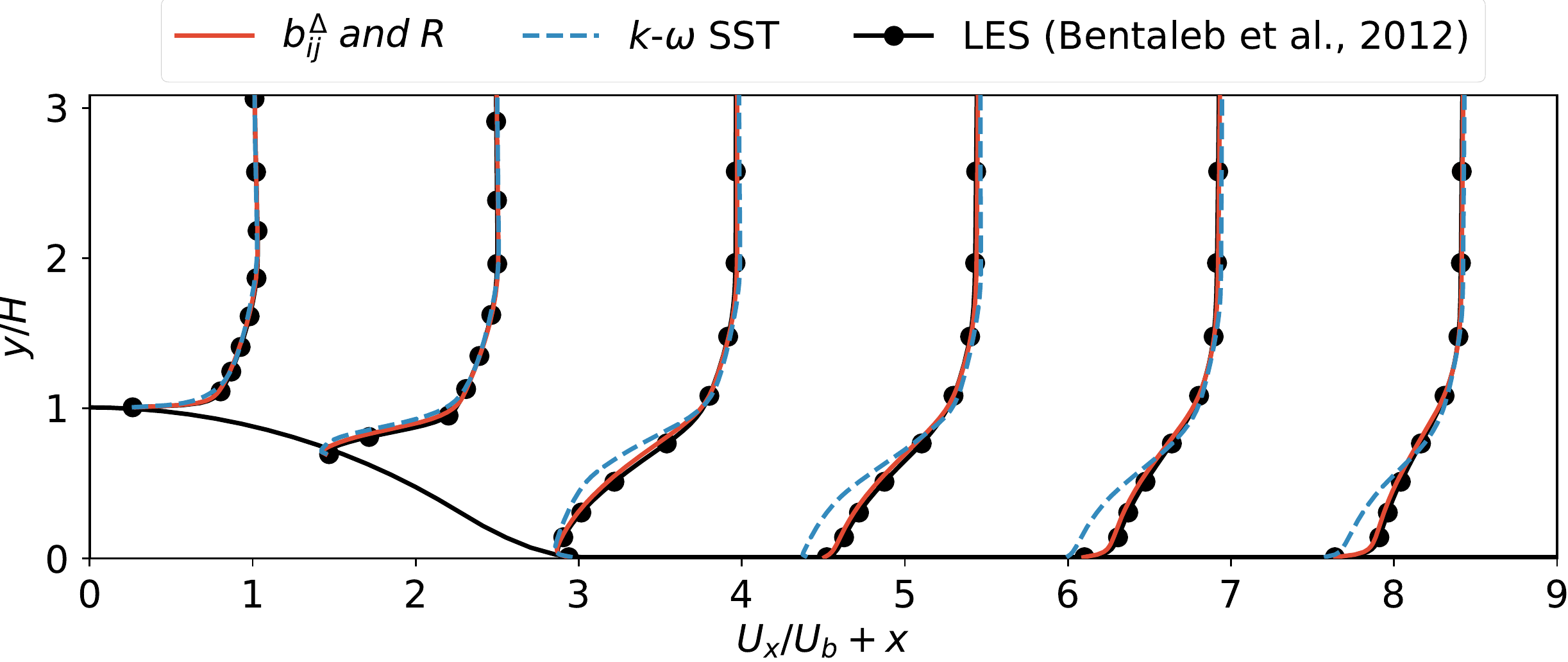}
    }
    \caption{Stream-wise velocity component for propagated model-form error acquired using \frozen{}.}
    \label{fig:verify}
\end{figure}

To evaluate $\bdelta$ it is necessary to estimate $\nu_t$, therefore also $\omega$ needs to be specified.  In \cite{Weatheritt2017b,Weatheritt2019}, $\omega$ was efficiently obtained by passively solving the $\omega$ transport equation given high-fidelity data for $U_i$, $k$ and $\bij$. The associated $\nu_t$ was then used to compute $\bdelta$ with \eqref{eq::nonlinearconstrel}. This method is named frozen-RANS as only one equation is solved iteratively while the remaining variables are frozen. Despite the fact that $\bdelta$ also alters the production of turbulent kinetic energy $P_k$, it is not evident that solving the $k$ equation given the data and the frozen $\omega$ should lead to the same $k$ as present in the data. Therefore, we introduce \frozen{} for which we also compute the residual of the $k$ equation alongside the computation of the frozen $\omega$. The residual is equivalent to an additive correction term, which we define as $R$, leading to an augmented \komegasst{} model
\begin{align}
	\partial_t k + U_j \partial_j k &= P_k + R - \beta^* \omega k + \partial_j \left[ (\nu + \sigma_k \nu_t) \partial_j k \right],\label{eq::augmentedkeq} \\
	\partial_t \omega + U_j \partial_j \omega &= \frac{\gamma}{\nu_t} \left( P_k + R\right) - \beta \omega^2 + \partial_j \left[ (\nu + \sigma_\omega \nu_t) \partial_j \omega \right] + CD_{k\omega},
	 \label{eq::augmentedkOmegaSST}
\end{align}
\noindent in which the production of turbulent kinetic energy is augmented by $\bdelta$ to $P_k = 2k(\bij^o + \bdelta) \partial_j U_i$. The corresponding eddy viscosity is $\nu_t = \frac{a_1 k}{\text{max}(a_1 \omega, SF_2)}$. The other standard terms of \komegasst{} read
\begin{align}	
	 CD_{k\omega} &= \max \left( 2 \sigma_{\omega 2} \frac{1}{\omega}(\partial_i k)(\partial_i \omega), 10^{-10} \right), \nonumber\\
	 F_1 &= \text{tanh}\left[\left(\min \left[\max \left(\frac{\sqrt{k}}{\beta^* \omega y}, \frac{500\nu}{y^2\omega} \right), \frac{4\sigma_{\omega 2}k}{CD_{k\omega}y^2} \right] \right)^4\right],\nonumber\\
	 F_2 &= \tanh \left[ \left( \max \left( \frac{2\sqrt{k}}{\beta^* \omega y}, \frac{500\nu}{y^2 \omega} \right) \right)^2  \right], \nonumber\\
	 \Phi &= F_1\Phi_1 + (1-F_1)\Phi_2,\nonumber\\
\end{align}
\noindent in which the latter blends the coefficients $\Phi \rightarrow (\Phi_1, \Phi_2)$
\begin{align}
	\alpha &= (5/9,0.44), \beta = (3/40,0.0828), \sigma_k = (0.85,1.0), \sigma_\omega = (0.5,0.856).
\end{align}
The remaining terms are 
 $\beta^*=0.09, a_1 = 0.31$ and $S = \sqrt{2S_{ij}S_{ij}}$. During the iterative computation of the frozen $\omega$ the residual of the $k$ equation is fed back into the $\omega$ equation until convergence is achieved. In order to validate that the resulting fields compensate the model-form error, $\bdelta$ and $R$ are added as static fields to a modified OpenFOAM solver \cite{Weller1998} and a CFD simulation is performed starting from the baseline solution for the flow configurations described in Section \ref{sec:testcases}, for which high-quality data is available. The mean-squared error between the high-fidelity data and the reconstructed velocity $U_i$ as well as the Reynolds-stress $\tau_{ij}$ is low, see Table \ref{tab:verify:mse}. Also the stream-wise velocity profiles shown in Figure \ref{fig:verify} demonstrate that the high-fidelity mean-flow data is essentially reproduced given $\bdelta$ and $R$. The \frozen{} approach requires full-field data, but is not based on an inversion procedure, e.g.~using adjoint-based optimisation as in \cite{Parish2016,Singh2017a}, which makes it very cost-efficient.
 \begin{table}
\caption{Mean-squared error $\epsilon$ of reconstructed velocity $U_i$ and Reynolds-stress $\tau_{ij}$ for different test cases with $\bdelta$ and $R$ added as static fields to the solver. Normalisation with $\epsilon$ of the baseline \komegasst{} results $U^o_i$ and $\tau^o_{ij}$. Description of cases in Section \ref{sec:testcases}.}
\label{tab:verify:mse}       
\begin{tabular}{lllll}
\hline\noalign{\smallskip}
Case & $\epsilon(U_i)\cdot10^{-5}$ & $\epsilon(U_i)/\epsilon(U^o_i)$ & $\epsilon(\tau_{ij})\cdot10^{-6}$ & $\epsilon(\tau_{ij})/\epsilon(\tau^o_{ij})$ \\
\noalign{\smallskip}\hline\noalign{\smallskip}
\ph{}   & $1.74$ & $0.00165$ & $36.7$ & $0.1495$\\
\cd{}   & $31.4$ & $0.0229$  & $7.21$ & $0.4781$\\
\cbfs{} & $59.6$ & $0.22703$ & $1.34$ & $0.4949$\\
\noalign{\smallskip}\hline
\end{tabular}
\end{table}

\subsection{Nonlinear eddy-viscosity models for $\bdelta$ and $R$}
In order to discover corrections for the model-form error $\bdelta$ and $R$, we need to decide on a modelling ansatz. Within this mathematical framework the symbolic regression targets to find specific expressions as corrections models. In \cite{Pope1975}, a nonlinear generalisation of the linear eddy viscosity concept was proposed. This concept has been used in several works on data-driven turbulence modelling \cite{Xiao2019,Duraisamy2019}. The fundamental assumption is made that the anisotropy of the Reynolds-stress $\bij$ not only depends on the strain rate tensor $S_{ij} = \tau \frac{1}{2} (\partial_j U_i + \partial_i U_j)$ but also on the rotation rate tensor $\Omega_{ij}= \tau \frac{1}{2} (\partial_j U_i - \partial_i U_j) $ with the timescale $\tau=1/\omega$. The Cayley-Hamilton theorem then dictates that the most general form of the anisotropic part of the Reynolds-stress can be expressed as
\begin{align}
	\bij(\Sij, \Oij) = \sum_{n=1}^N T_{ij}^{(n)} \alpha_n(I_1, ..., I_5), \label{eq::nonlinearstressstrain}
\end{align}
\noindent with ten nonlinear base tensors $T_{ij}^{(n)}$ and five corresponding invariants $I_m$. Only the first four base tensors and the first two invariants are used in this work, which are
\begin{align}
	T_{ij}^{(1)} &= S_{ij},\nonumber \\ 
	T_{ij}^{(2)} &= S_{ik}\Omega_{kj} - \Omega_{ik}S_{kj}, \nonumber \\
	T_{ij}^{(3)} &= S_{ik}S_{kj} - \frac{1}{3} \delta_{ij} S_{mn}S_{nm}, \nonumber \\ 
	T_{ij}^{(4)} &= \Omega_{ik}\Omega_{kj} - \frac{1}{3} \delta_{ij} \Omega_{mn}\Omega_{nm} \label{eq:basetensor}
\end{align}
\begin{align}
	I_1 = S_{mn}S_{nm}, I_2 &= \Omega_{mn}\Omega_{nm}. \label{eq:invariants}
\end{align}
Using this set for \eqref{eq::nonlinearstressstrain} we have an ansatz, which only requires functional expressions for the coefficients $\alpha_n$, to model $\bdelta$. However, computing $\bdelta$ using \eqref{eq::nonlinearconstrel} requires a correct $k$ as discussed in Section \ref{subsec::additive_modelform_error}. This aspect is taken into account in the modelling ansatz for $R$, for which we take a closer look at the eddy viscosity concept.

Both linear and nonlinear eddy viscosity models provide expressions for the anisotropy $\bij$ based on a local relation between stress and strain. Due to the restriction of this local closure only the normal stresses $\frac{2}{3}k\delta_{ij}$ can account for nonlocal effects by transport equations for the turbulent quantities using convection and diffusion terms \cite{Leschziner2015,Wilcox}. The term $R$ provides local information to correct the transport equations. Depending on the local sign of $R$ it either increases or decreases the net production $P_k$ locally. Hence, it acts as an additional production or dissipation term, which can overcome the error in $k$. We model it in a similar way to the turbulent production
\begin{align}
	R &= 2k\bpk \partial_j U_i,
\end{align}
\noindent which has the additional benefit that we can also use the framework of nonlinear eddy viscosity models to model $R$. 

Since the general modelling framework is the same for both $\bdelta$ and $R$, a natural next step would be to combine both in order to find a single model accounting for the sources of model-form error on the level of the constitutive relation as well as within the turbulent transport equations. For example in \cite{Weatheritt2017b} models identified using genetic programming were modified such that any additional contribution of the first base tensor $T^{(1)}_{ij}$ in \eqref{eq::nonlinearstressstrain} was added with a positive sign for the computation of $P_k$. This ad-hoc correction was established based on physical reasoning to avoid very low production close to walls and led to significantly improved predictions. However, in contrast to \cite{Weatheritt2017b} we have extracted two target terms $\bdelta$ and $R$ using \frozen{}, which also make it possible to systematically study (i) how to obtain corrections models for each target individually and (ii) their combined effect on the predictions. 

Given the polynomial model \eqref{eq::nonlinearstressstrain} and the set of base tensors \eqref{eq:basetensor} and invariants \eqref{eq:invariants} we are now left with the task of providing suitable expressions for $\alpha_n(I_1, I_2)$ for $n=1, ..., 4$ to overcome the model-form error. This is the purpose of the deterministic symbolic regression technique detailed in the following section.

\section{Model discovery methodology}
\label{sec:symbolicregression}
Deterministic symbolic regression constructs a large library of nonlinear candidate functions to regress data. It identifies the relevant candidates by adopting a sparsity constraint. Two fundamental methods have been proposed: Sparse identification of nonlinear dynamics (SINDy) \cite{Brunton2016,Brunton2019} and fast function extraction (FFX) \cite{McConaghy2011}. Both methods were applied in several areas of physical modelling. In the following, we introduce the steps of the model discovery methodology \method{} based on FFX, for which a library is constructed using a set of raw input variables and mathematical operations. The model selection uses elastic net regression. Finally, for the inference of the model coefficients the stability requirements of a CFD solver are considered. An overview of \method{} is given in Figure \ref{fig:diagram}.

\subsection{Building a library of candidate functions}
The deterministic symbolic regression requires a library of candidate functions, from which a model is deduced by building a linear combination of the candidates. Hence, the library is an essential element of the entire methodology and needs to accommodate relevant candidates explaining the data. We rely on the nonlinear eddy viscosity concept and aim to find models for $\alpha_n$ in \eqref{eq::nonlinearstressstrain} given as primitive input features the invariants $I_1$ and $I_2$. For the present work we focus on a library, in which the primitive input features are squared and the resulting candidates are multiplied by each other leading to a maximum degree of 6. In addition to the two invariants we also include a constant function $c$ to the set of raw input features. The resulting vector $\B$ reads
\begin{align}
	\B =& \Big[ c, I_1, I_2, I_1^2, I_2^2, I_1^2 I_2^3, I_1^4 I_2^2, I_1 I_2^2, I_1 I_2^3,\nonumber \\ 
	& \;\;I_1 I_2^4, I_1^3 I_2, I_1^2 I_2^4, I_1^2 I_2, I_1 I_2, I_1^3 I_2^2, I_1^2 I_2^2 \Big]^T
	\end{align}
\noindent with the cardinality of $\B$, $|\B|=16$.

For the library to regress models for $\bdelta$ each function of $\B$ is multiplied with each base tensor $T_{ij}^{(n)}$, leading to the library of tensorial candidate functions
\begin{align}
	\C{\bdelta} = \left[c\T{1}, c\T{2}, \hdots, I_1^2 I_2^2\T{4} \right]^T.
	\label{eq:library:C:bdelta}
\end{align}
In order to regress models for $R$ the double dot product of each function in $\C{\bdelta}$ with the mean velocity gradient tensor $\partial_j U_i$ is computed, leading to
\begin{align}
	\C{R} = \left[c\T{1}\partial_j U_i,  \hdots, I_1^2 I_2^2\T{4}\partial_j U_i \right]^T.
	\label{eq:library:C:Pd}
\end{align}
The two libraries $\C{\bdelta}$ and $\C{R}$ are evaluated given the high-fidelity validation data for each test case and stored column-wise in matrices
\begin{align}
	\Cd{\bdelta} &= 
	\begin{bmatrix}
		c T_{xx}^{(1)}|_{k=0} & c T_{xx}^{(2)}|_{k=0} & \hdots & I_1^2 I_2^2 T_{xx}^{(4)}|_{k=0}\\
		c T_{xy}^{(1)}|_{k=0} & c T_{xy}^{(2)}|_{k=0} & \hdots & I_1^2 I_2^2 T_{xy}^{(4)}|_{k=0}\\
		c T_{xz}^{(1)}|_{k=0} & c T_{xz}^{(2)}|_{k=0} & \hdots & I_1^2 I_2^2 T_{xz}^{(4)}|_{k=0}\\
		c T_{yy}^{(1)}|_{k=0} & c T_{yy}^{(2)}|_{k=0} & \hdots & I_1^2 I_2^2 T_{yy}^{(4)}|_{k=0}\\		
		c T_{xz}^{(1)}|_{k=0} & cT_{yz}^{(2)}|_{k=0} & \hdots & I_1^2 I_2^2 T_{yz}^{(4)}|_{k=0}\\		
		c T_{zz}^{(1)}|_{k=0} & c T_{zz}^{(2)}|_{k=0} & \hdots & I_1^2 I_2^2 T_{zz}^{(4)}|_{k=0}\\		
		\vdots & \vdots &   & \vdots\\
		c T_{zz}^{(1)}|_{k=K} & c T_{zz}^{(2)}|_{k=K} & \hdots & I_1^2 I_2^2 T_{zz}^{(4)}|_{k=K}\\
    \end{bmatrix} \in \mathbb{R}^{6K \times |\C{\bdelta}|},\label{eq:Cd:bdelta}\\
    \Cd{R} &= 
	\begin{bmatrix}
		c T_{ij}^{(1)} \partial_j U_i|_{k=0} & \hdots & I_1^2 I_2^2 T_{ij}^{(4)} \partial_j U_i|_{k=0}\\		
		\vdots &    & \vdots\\
		c T_{ij}^{(1)} \partial_j U_i|_{k=K} & \hdots & I_1^2 I_2^2 T_{ij}^{(4)} \partial_j U_i|_{k=K}\\
    \end{bmatrix}\in \mathbb{R}^{K \times |\C{R}|},\label{eq:Cd:R}
\end{align}
\noindent in which $K$ is the number of mesh points of the test case at hand. The corresponding target data $\bdelta$ and $R$ are stacked to vectors
\begin{align}
\bmbdelta &= \left[ b^\Delta_{xx}|_{k=0}, b^\Delta_{xy}|_{k=0}, ... , b^\Delta_{zz}|_{k=K} \right]^T \in \mathbb{R}^{6K}, \\
\bm{R} &= \left[ R|_{k=0}, R|_{k=1}, ... , R|_{k=K} \right]^T \in \mathbb{R}^{K}.
\end{align}

\subsection{Model selection using sparsity-promoting regression}
\label{subsec:symbolicregression}
Given the above defined libraries the task is to form a linear model to regress the target data $\bm{\Delta}=\bmbdelta$ or $\bmR$ by finding the coefficient vector $\bmtheta$
\begin{align}
	\bm{\Delta} = \Cd{\Delta} \bmtheta,
\end{align}
\noindent which represents a large, overdetermined system of equations. When using ordinary least-squares regression a dense coefficient vector $\bmtheta$ is obtained, resulting in overly complex models, which are potentially overfitting the data given the large libraries  \eqref{eq:library:C:bdelta} and \eqref{eq:library:C:Pd}. Due to multi-collinearity between the candidates, $\Cd{\Delta}$ can be ill-conditioned, so that the coefficients may also display large differences in magnitude expressed in a large $l_1$-norm of $\bmtheta$. Such models are unsuitable to be implemented in a CFD solver as they increase the numerical stiffness of the problem and impede convergence of the solution. 

Following the idea of parsimonious models we constrain the search to models which optimally balance error and complexity and are not overfitting the data \cite{Brunton2019}. In principle, given a library a combinatoric study can be carried out, by performing an ordinary least-squares regression for each possible subset of candidates. Starting from each single candidate function individually, proceeding with all possible pairs up to more complex combinations. As the number of possible models grows exponentially with the number of candidates $I = 2 ^ {|\C{\Delta}|} -2$ this approach becomes already infeasible for the simple libraries \eqref{eq:library:C:bdelta} and \eqref{eq:library:C:Pd} with $|\C{\Delta}| \approx 64$.

Hence, we follow \cite{Brunton2019,McConaghy2011} and engage sparsity-promoting regularisation of the underlying least-squares optimisation problem. The model-discovery procedure is divided into two parts: (i) model selection and (ii) model inference, see Figure \ref{fig:diagram}. For the first step, the model selection, we use the elastic net formulation
\begin{align}
	\bm{\Theta} = \argmin_{\hat{\bm{\Theta}}} \norm{ \Cd{\Delta} \hat{\bm{\Theta}} - \bm{\Delta}}_2^2 + \lambda \rho \norm{\hat{\bm{\Theta}}}_1 \nonumber \\
		+ \,0.5 \lambda  (1-\rho) \norm{\hat{\bm{\Theta}}}_2^2,
		\label{eq:opt}
\end{align}
\noindent which blends the $l_1$- and $l_2$-norm regularisation given the mixing parameter $\rho \in [0,1]$ and the regularisation weight $\lambda$, to promote the sparsity of $\bm{\Theta}$ \cite{McConaghy2011,Zou2005}. On its own, the $l_1$-norm, known as Lasso-regression, promotes sparsity by allowing only a few nonzero coefficients while shrinking the rest to zero. The $l_2$-norm, known as Ridge-regression, enforces relatively small coefficients without setting them to zero, but is able to identify also correlated candidate functions instead of picking a single one. By combining both methods, the elastic net can find sparse models with a good predictive performance. Besides the mixing parameter, also the regularisation parameter $\lambda$ shapes the form of the model: For a very large $\lambda$ the vector $\bm{\Theta}$ will only contain zeros independent of $\rho$. The amount of nonzero coefficients increases for smaller $\lambda$ values making the discovery of sparse models possible. 

Given the elastic net regularisation method we need to specify suitable combinations of the weight $\lambda$ and type of the regularisation $\rho$, for which the optimisation problem \eqref{eq:opt} is solved. Most commonly the optimal $(\lambda, \rho)$ combination is found based on a strategy to avoid overfitting of the resulting models, e.g. using cross-validation \cite{Brunton2019}, for which the data is split into a training and a test set. While the optimisation problem given a grid $(\bm{\lambda}, \bm{\rho})$ is solved on the former, only the model with the best performance evaluated on the latter survives. For the purpose of CFD a true validation of the models can only be performed once they are implemented in a solver and applied to a test case. In order to not overcharge the role of the training data from \frozen{} at this stage of the methodology, we select a wide spectrum of models varying in accuracy and complexity using \eqref{eq:opt} instead of a single one. The validation task will be performed later using a CFD solver.

Following \cite{McConaghy2011} we use
\begin{align}
	\bm{\rho} &= [0.01, 0.1, 0.2, 0.5, 0.7, 0.9, 0.95, 0.99, 1.0]^T,
\end{align}
\noindent which ensures that we cover a substantial range of different regularisation types. The upper limit of the regularisation weight is defined as $\lambda_{\text{max}} = \max(|\Cd{\Delta}^T \bm{\Delta}|) / (K \rho)$, because for any $\lambda>\lambda_{\text{max}}$ all elements in $\bmtheta$ will be equal to zero. The entire vector 
\begin{align}
	\bm{\lambda} &= [\lambda_{0}, ..., \lambda_{\text{max}}]^T
\end{align}
\noindent is defined of having $100$ entries between $\lambda_{0} = \xi \lambda_{\text{max}}$ with $\xi=10^{-3}$ uniformly spaced using a log-scale as defined in  \cite{McConaghy2011}. This provides a search space $(\bm{\lambda}, \bm{\rho})$, the elastic net, which is large enough and has an appropriate resolution. At each grid point $(\lambda_{i}, \rho_j)$ a vector $\bmtheta^{(i,j)}_\Delta$  as a solution of \eqref{eq:opt} is found using the coordinate descent algorithm. The duration for the model selection step given the number of data points $K \sim 15000 $ is of the order of a minute on a standard consumer laptop.

Solving \eqref{eq:opt} for different $(\lambda_{i}, \rho_j)$ might produce $\bmtheta^{(i,j)}_\Delta$ with the same abstract model form $\bm{\bar{\Theta}}$, which means that the same entries are equal to zero. As the specific values of the coefficients will be defined in the next step, the selection step of \method{} concludes with filtering out the set of $D$ unique abstract model forms $\bm{\mathcal{D}}_{\Delta} = \left\{\bm{\bar{\Theta}}_\Delta^d \big| d = {1, ..., D} \right\}$. 

\subsection{Model inference for CFD}
\label{sec:inference}
The abstract models $\bm{\mathcal{D}}_{\Delta}$ are found using standardised candidates, because the relevance of each candidate should not be determined by its magnitude during the model selection step. With the aim of defining a model with the correct units, we need to perform an additional regression using the unstandardised candidate functions for each subset determined by the abstract model forms in $\bm{\mathcal{D}}_{\Delta}$, which is the purpose of the model inference step outlined in the following. 

In \cite{Brunton2019,Quade2018,Mangan2017} this was done using ordinary least-squares regression for problems in the domains of dynamical systems and biological networks. As mentioned above, the ability of the CFD solver, in which the models will be implemented, to produce a converged solution is sensitive to large coefficients, which has been reported in \cite{Weatheritt2016,Weatheritt2017b,Weatheritt2019}. We take this additional constraint into account by performing a Ridge regression
\begin{align}
	\bm{\Theta}_\Delta^{s,d} = \argmin_{\hat{\bm{\Theta}}_\Delta^{s,d}} \norm{ \Cd{\Delta}^s \hat{\bm{\Theta}}_\Delta^{s,d} - \bm{\Delta}}_2^2 + \, \lambda_r   \norm{\hat{\bm{\Theta}}_\Delta^{s,d}}_2^2,
		\label{eq:ridge}
\end{align}
\noindent in which $\lambda_r$ is the Tikhonov-regularisation parameter. The index $s$ denotes the submatrix of $\Cd{\Delta}$ and the subvector of $\bm{\Theta}_\Delta^{d}$ consisting of the selected columns or elements respectively as defined in $\bm{\mathcal{D}}_{\Delta}$. The elements of $\bm{\Theta}_\Delta^d$ associated with the inactive candidates are zero and are not modified during this step. 

By using the $l_2$-norm regularisation the magnitude of the nonzero coefficients is shrunk \cite{Brunton2019,Bishop2006}. In general, low values for $\lambda_r$ reduce the bias introduced through regularisation, but lead to larger coefficient values, and vice versa. Since shrinkage of the coefficients also reduces the influence of candidate functions with a lower magnitude compared to others, we need to find a trade-off between error of the model on the target data $\bmdelta$ and the likelihood that the model will deliver converged solutions when used in a CFD solver. The problem of finding such an optimum is that the latter aspect can only be answered retrospectively. Recently, this problem has been addressed in \cite{Zhao2019} by embedding CFD simulations in the search for correction models guided by genetic programming. While this increases the costs of the model search drastically, it also significantly increases the chance of delivering models with better convergence properties. Even though this procedure provides a strong indication, the identified models are also not guaranteed to converge \emph{a priori} for any other test case outside the training set. Via testing using the cases in Section \ref{sec:testcases}, we have identified $0.1<\lambda_r<0.01$ able to deliver coefficients in a range balancing the error on the target data $\Delta$ and the likelihood to produce converged CFD solutions. 

Our efforts are based on an empirical observation, but do not guarantee a well-behaving numerical setup under all conditions. However, we have identified corrections of $\bdelta$ as the only contribution which can do harm to the convergence properties for the given test cases. Therefore, if a model does not converge, we further decrease the coefficients by a factor $\xi=0.1$, for the model correcting $\bdelta$ only. This ad-hoc intervention is sufficient to achieve convergence for the studied cases. 

Finally, the resulting coefficient vector $\bm{\Theta}_\Delta^d$ is used to retrieve the symbolic expression of the models by a dot product with the library of candidate functions $\C{\Delta}$ in \eqref{eq:library:C:bdelta} and \eqref{eq:library:C:Pd}
\begin{align}
	M^d_\Delta := \C{\Delta}^T \bm{\Theta}_\Delta^d,
	\label{eq:mod}
\end{align}
\noindent which are implemented in the open-source finite-volume code OpenFOAM \cite{Weller1998}. The divergence terms of the equations are discretised with linear upwinding and turbulent diffusion with 2nd order central differencing. In summary, the model discovery step of \method{} selects models utilising elastic net regression in \eqref{eq:opt} and further infers the coefficients of the selected models in \eqref{eq:ridge}. The latter process is guided by the aim to discover models complying with the restrictions of a CFD solver.

\section{Test cases and high-fidelity data}
\label{sec:testcases}
In order to apply \method{} we need full-field data of $U_i$, $k$ and $\tau_{ij}$, which we take from LES and DNS studies conducted by other researchers. We have selected three test cases of separating flows over curved surfaces in two-dimensions with similar Reynolds-numbers. For each case fine meshes are selected, which ensure that the discretisation error is much smaller compared to the error due to turbulence modelling.

\textbf{Periodic hills (PH)}, for which the flow is over a series of hills in a channel. Initially proposed by \cite{Mellen2000} this case has been studied both experimentally as well as numerically in detail. We use LES data from \cite{Breuer2009} for $Re=10595$ (\ph) to apply \method{} and test the performance of the resulting models. In addition, we also use experimental data from \cite{Rapp2011} at a much larger $Re=37000$ (\phexp) in order to test the models outside the range of the training data. The numerical mesh consists of $120 \times 130$ cells. Cyclic boundary conditions are used at the inlet and outlet. The flow is driven by a volume forcing defined to produce a constant bulk velocity.

\textbf{Converging-diverging channel (CD)}. A DNS study of the flow within a channel, in which an asymmetric bump is placed, exposed to an adverse pressure gradient was performed by \cite{Laval2011} for $Re=12600$ (\cd). The flow shows a small separation bubble on the lee-side of the bump, which is challenging for RANS to predict. The numerical mesh consists of $140 \times 100$ cells. The inlet profile was obtained from a channel-flow simulation at equivalent $Re$.

\textbf{Curved backward-facing step (CBFS)}. In \cite{Bentaleb2012} a LES simulation of a flow over a gently-curved backward-facing step was performed at $Re=13700$ (\cbfs). Similar to PH also for this flow the mean effect of separation and reattachment dynamics is the objective. The numerical mesh consists of $140 \times 150$ cells. The inlet was obtained from a fully-developed boundary layer simulation.

Despite the simple geometries, the mean effect of the separation and reattachment dynamics of a flow on a curved surface is a challenging problem for steady-RANS approaches. Especially, PH serves as an important testbed for classical and data-driven approaches for turbulence modelling, e.g.~\cite{Xiao2019,Jakirlic2012}, but also the other two have been introduced with the purpose of closure investigation.

\section{Results}
The method \method{} introduced in Section \ref{sec:symbolicregression} is applied to the three test cases of Section \ref{sec:testcases}. The models resulting from the model-discovery are presented and their mean-squared error on the training data is evaluated. In order to identify the models with the best predictive capabilities, we carry out cross-validation of the resulting models using CFD \cite{Bishop2006}: Models identified given training data of one case are used for CFD simulations of the remaining two case. For each case a single model is chosen as the best-performing one. Finally, the three resulting models are tested in a \emph{true} prediction for the flow over periodic hills at $Re=37000$.
\subsection{Discovery of models and their error on training data}
\begin{figure}
	\subfloat[$\bdelta$ \label{fig:modelstruc:adelta:prior:ph} ]{%
        \includegraphics[width=\textwidth]{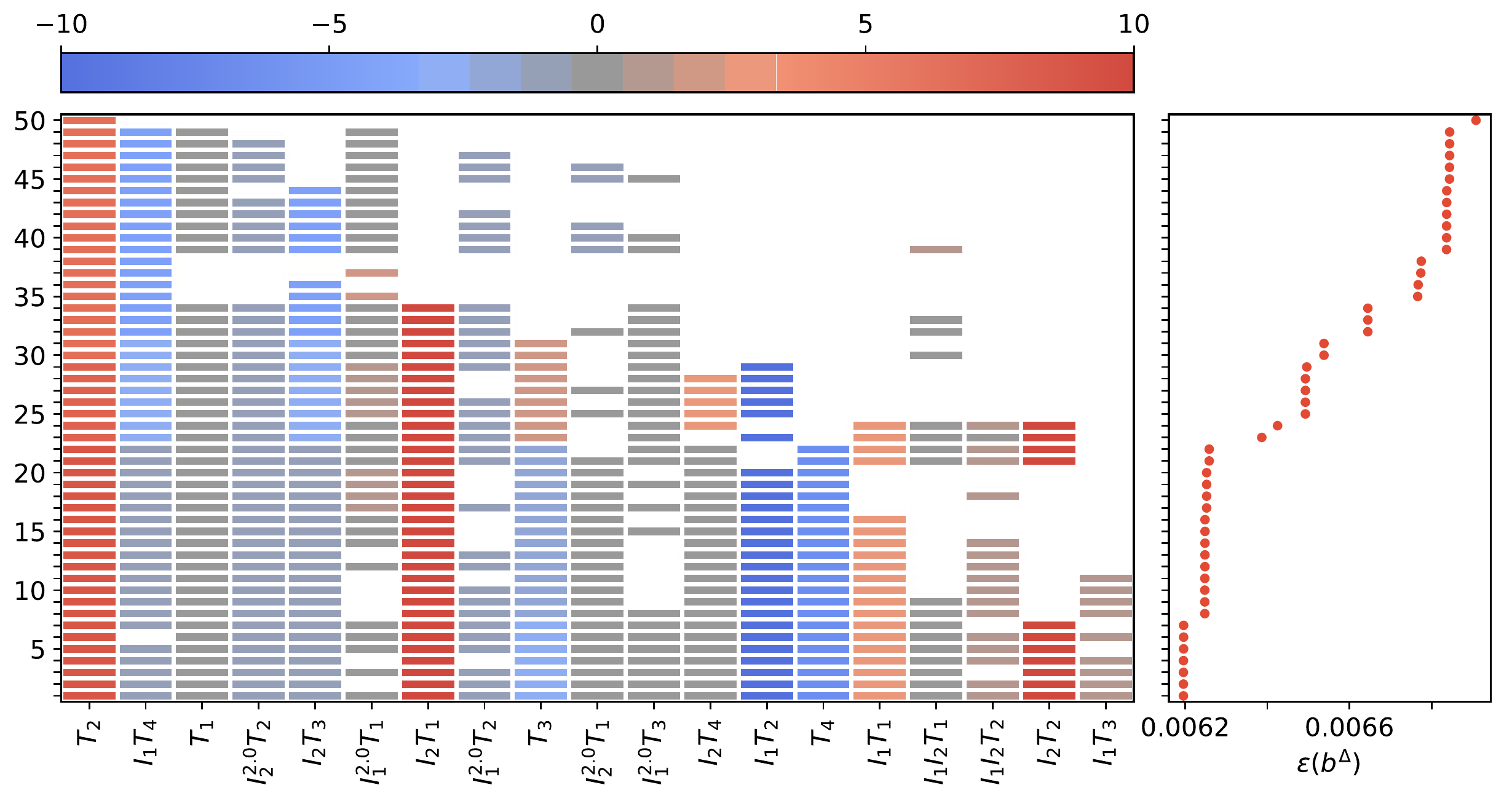}
    }
    \hfill
    \subfloat[$R$ \label{fig:modelstruc:R:prior:ph}]{%
		\includegraphics[width=\textwidth]{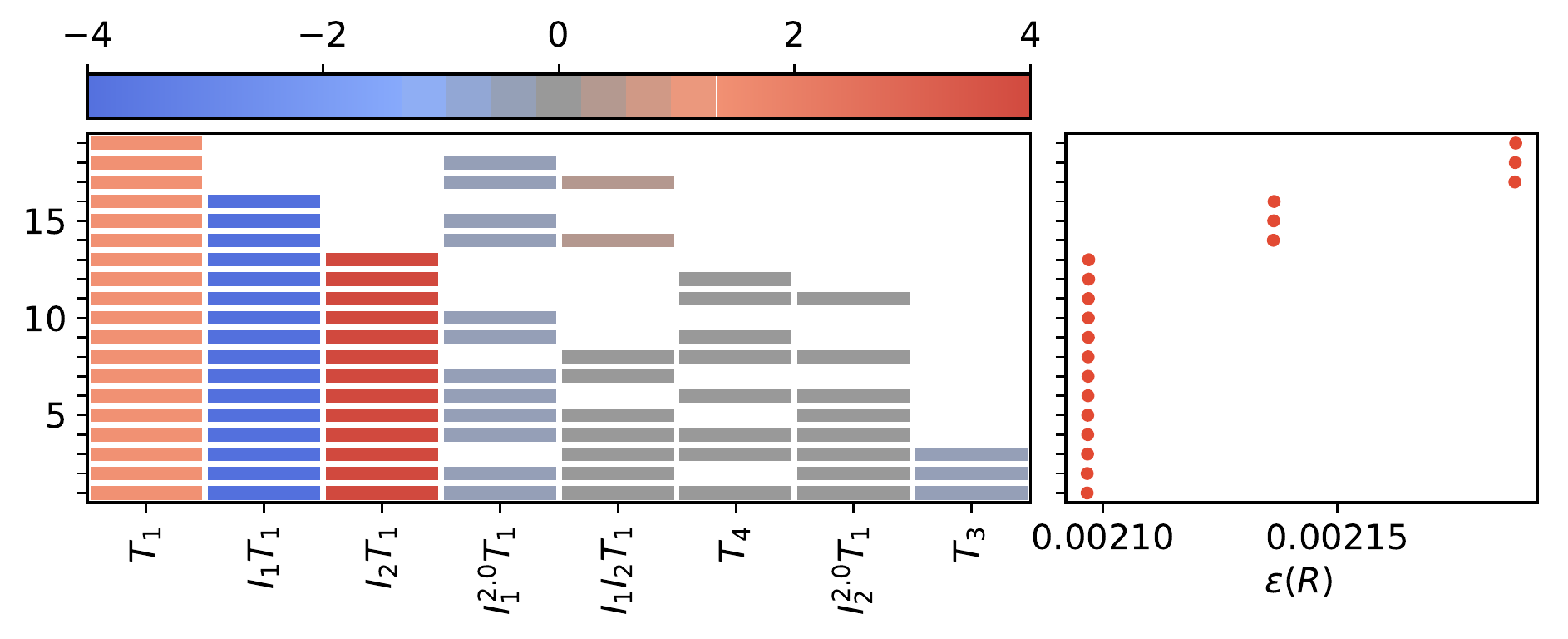}
    }
    \caption{Model-structure of all discovered models using \method{} and mean-squared error on training data for \ph. The matrix (l.) shows the values of the active (coloured) candidate functions (x-axis) for each model $M_i$ with model index $i$ (y-axis). The mean-squared error between the frozen data $\bdelta$ and the model is also shown (r.).}
    \label{fig:modelstruc:prior:ph}
\end{figure}

The goal of the model-discovery is to identify an ensemble of diverse models with small coefficients, varying in model-structure (complexity) and accuracy. Such an ensemble is better-suited for the cross-validation on unseen test cases, than a selection of the best models given only the training data. The sparse-regression for $\bdelta$ applied to the three test cases \ph{}, \cbfs{} and \cd{} resulted in 52, 114 and 136 distinct models respectively, see Figure \ref{fig:modelstruc:adelta:prior:ph} for the results for \ph. It can be observed that, in general, an increase in complexity of a model leads to a reduction of the error. But, the bias introduced through the ridge regression of the inference step in \method, see Section \ref{sec:inference}, shrinks the model coefficients. If a coefficient is associated with a candidate function with a much lower magnitude compared to others, due to shrunk coefficients it becomes less relevant. The result is a staircase structure of the error: Models show a different form but have similar error. This pattern can also be observed for the other cases and becomes even more prominent for the models regressing $R$. For this target the model discovery resulted in 18 and 19 distinct model forms for \cd{} and \ph{} respectively, see Figure \ref{fig:modelstruc:R:prior:ph} for results of case \ph. For \cbfs{} only three models have been found. We identify $\T{1}$, $I_1\T{1}$ and $I_2 \T{1}$ as the relevant candidates to regress $R$, and models combining all three give the lowest error per test case. 

\begin{figure}
	\subfloat[\ph{} \label{fig:modelstruc:Pd:prior:ph:all}]{%
       \includegraphics[width=\textwidth]{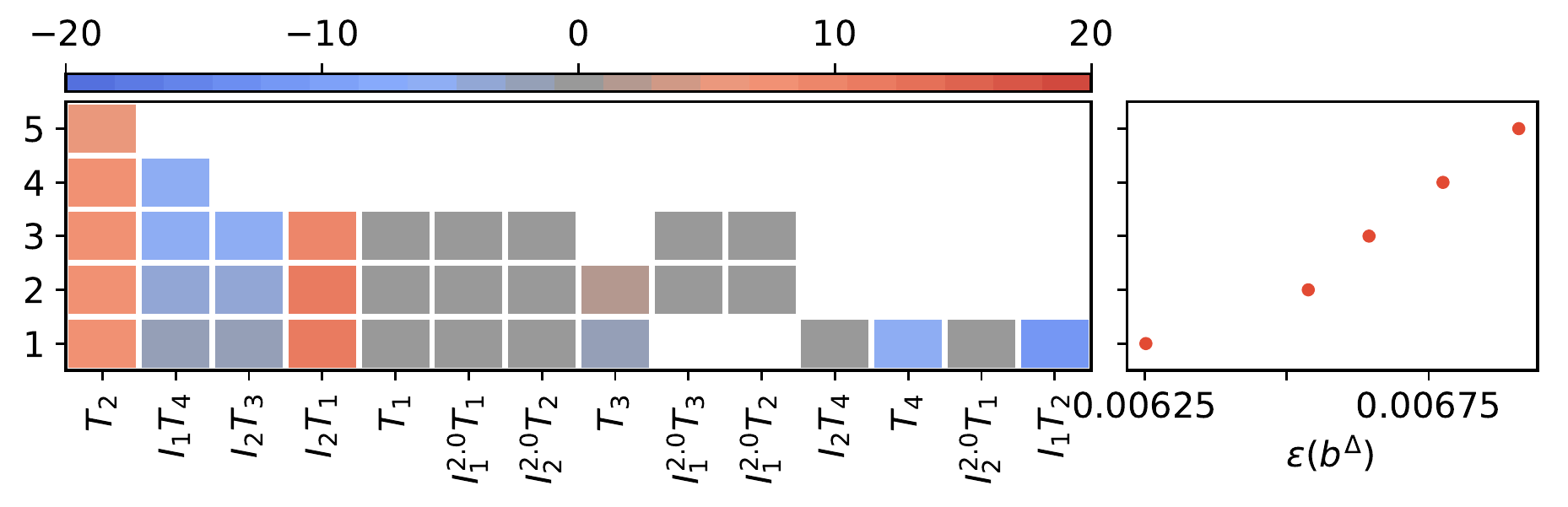}
    }
    \hfill
    \subfloat[\cd{} \label{fig:modelstruc:Pd:prior:ph:sel}]{%
		\includegraphics[width=\textwidth]{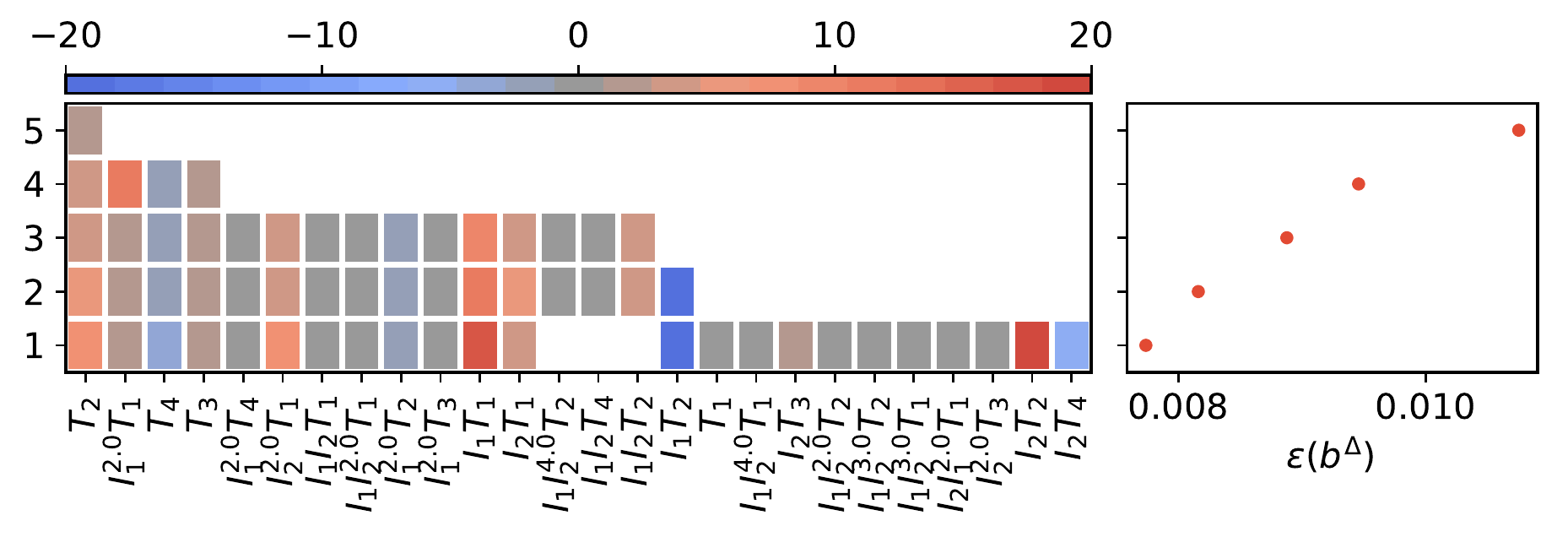}
    }
    \hfill
    \subfloat[\cbfs{} \label{fig:modelstruc:Pd:prior:ph:sel}]{%
		\includegraphics[width=\textwidth]{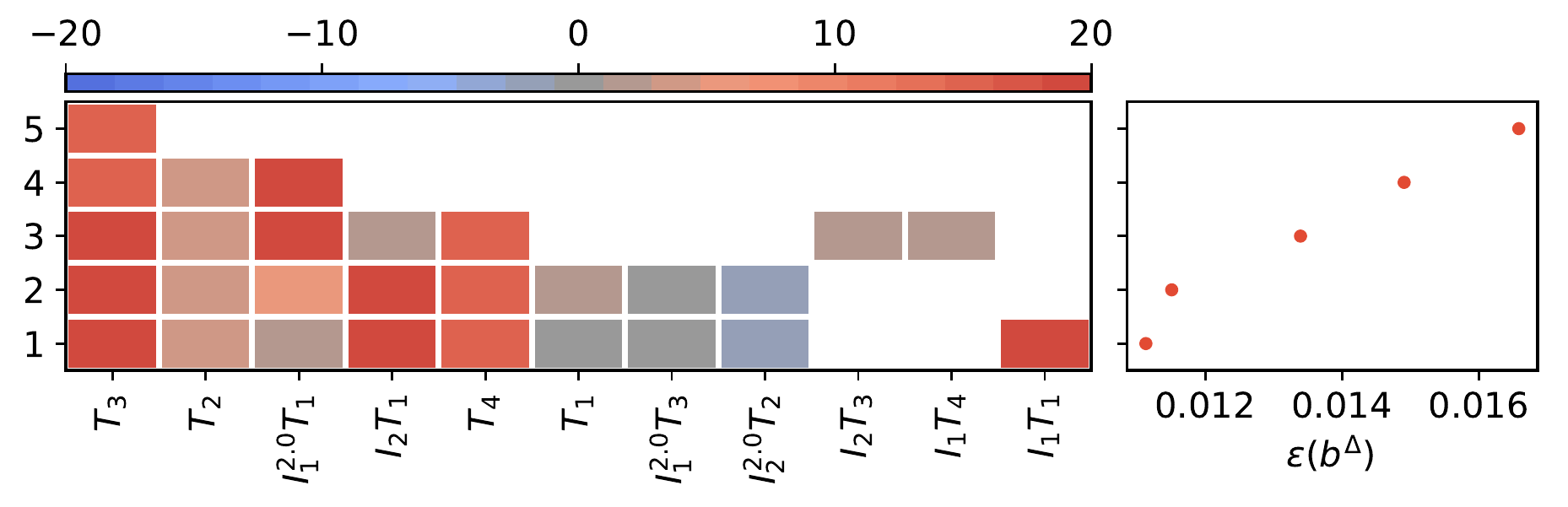}
    }
	    \caption{Selected models on frozen data $\bdelta$.}
    \label{fig:modelstruc:adelta:prior}
\end{figure}

\begin{figure}
  \includegraphics[width=\textwidth]{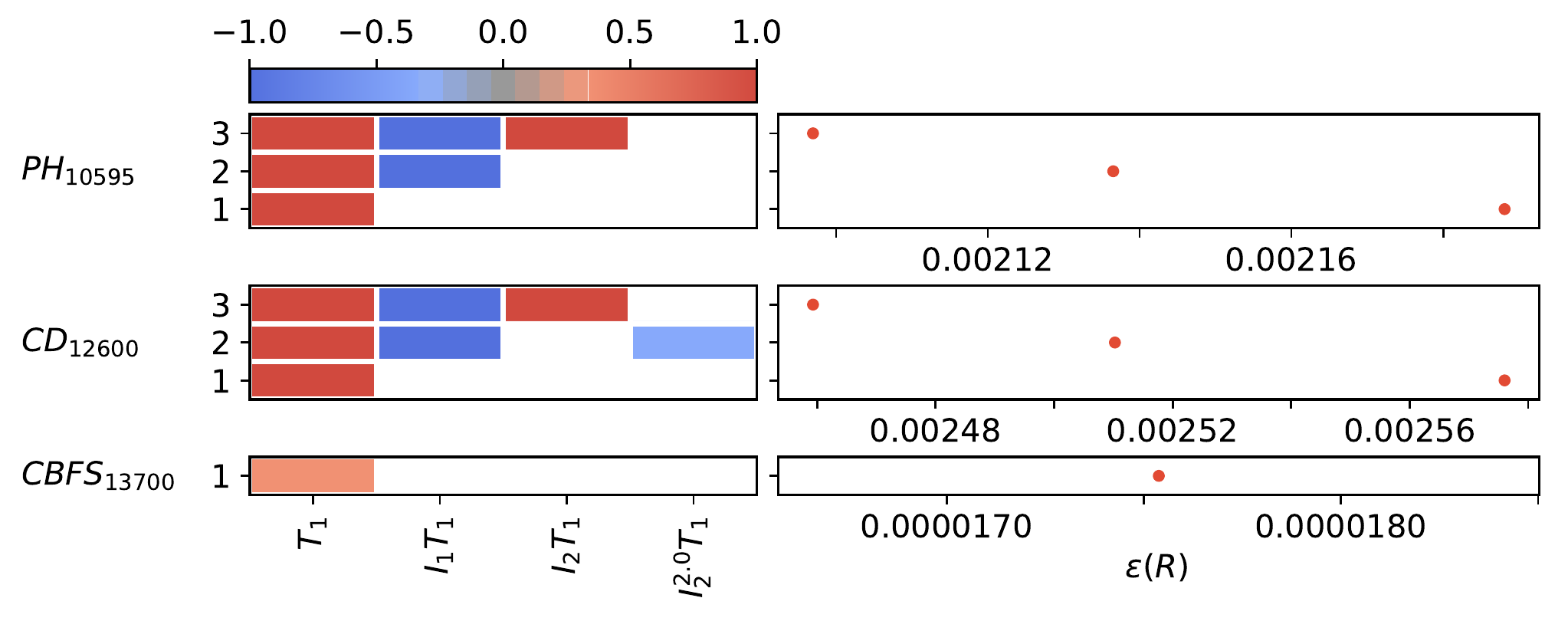}
\caption{Selected models on frozen data $R$.}
\label{fig:modelstruc:R:prior}       
\end{figure}

In order to reduce the redundancy within the ensemble of models regressing $\bdelta$ and $R$ we select only a representative  subset of models. This ensemble needs to acknowledge the hierarchical structure of diverse model-forms and their accuracy. In an ad-hoc way, we hand-select 5 models for $\bdelta$ and 3 for $R$, except for \cbfs{} only 1. The ensembles of selected models are shown in Figure \ref{fig:modelstruc:adelta:prior} and Figure \ref{fig:modelstruc:R:prior}. The result is a hierarchical spectrum of models regressing the training data varying in complexity and error. Given this ensemble we study the performance of each model for predictions in the next section.
\begin{figure}
  \includegraphics[width=\textwidth]{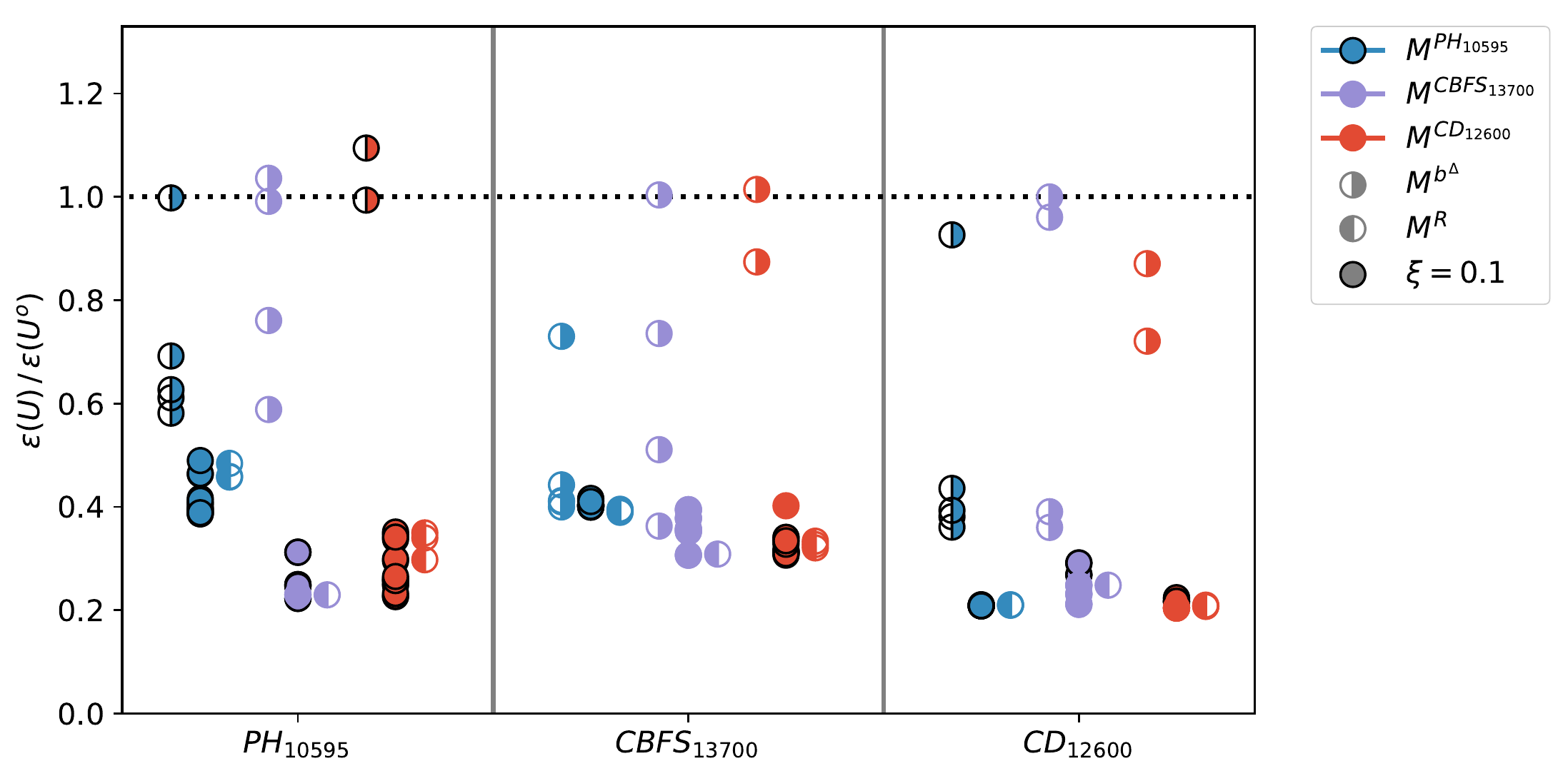}
\caption{Mean-squared error of velocity vector of each correction model normalised by the mean-squared error of the baseline \komegasst. The colour indicates on which high-fidelity data the models have been identified. Full circles represent simulations using both corrections, while left-/right-filled circles represent simulations using only correction for $R$ or $\bdelta$ respectively.}
\label{fig:U:MSE}       
\end{figure}%
\begin{figure}
	\subfloat[\ph{} \label{fig:modelstruc:ph}]{%
       \includegraphics[width=\textwidth]{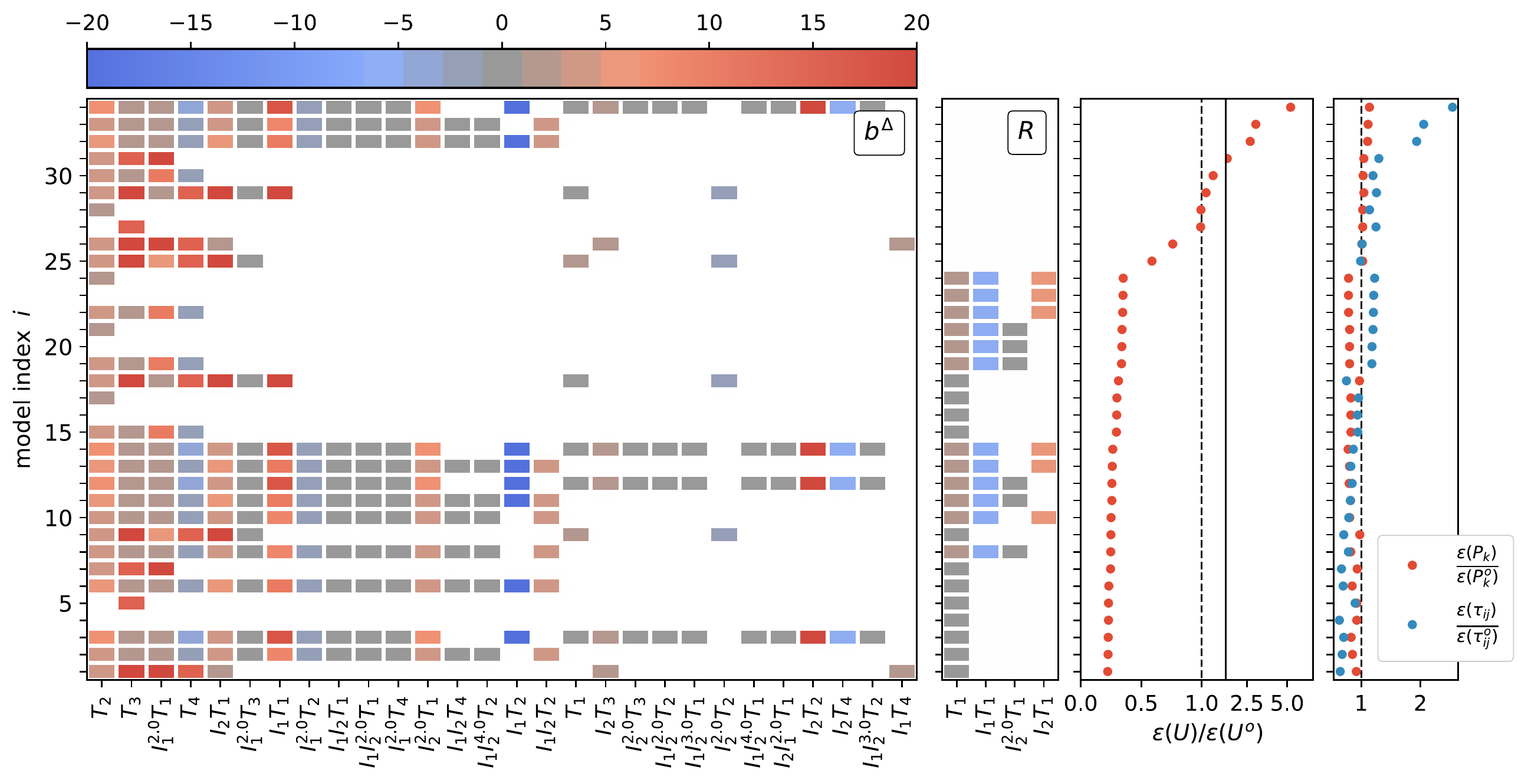}
    }
    \hfill
    \subfloat[\cd{} \label{fig:modelstruc:cd}]{%
		\includegraphics[width=\textwidth]{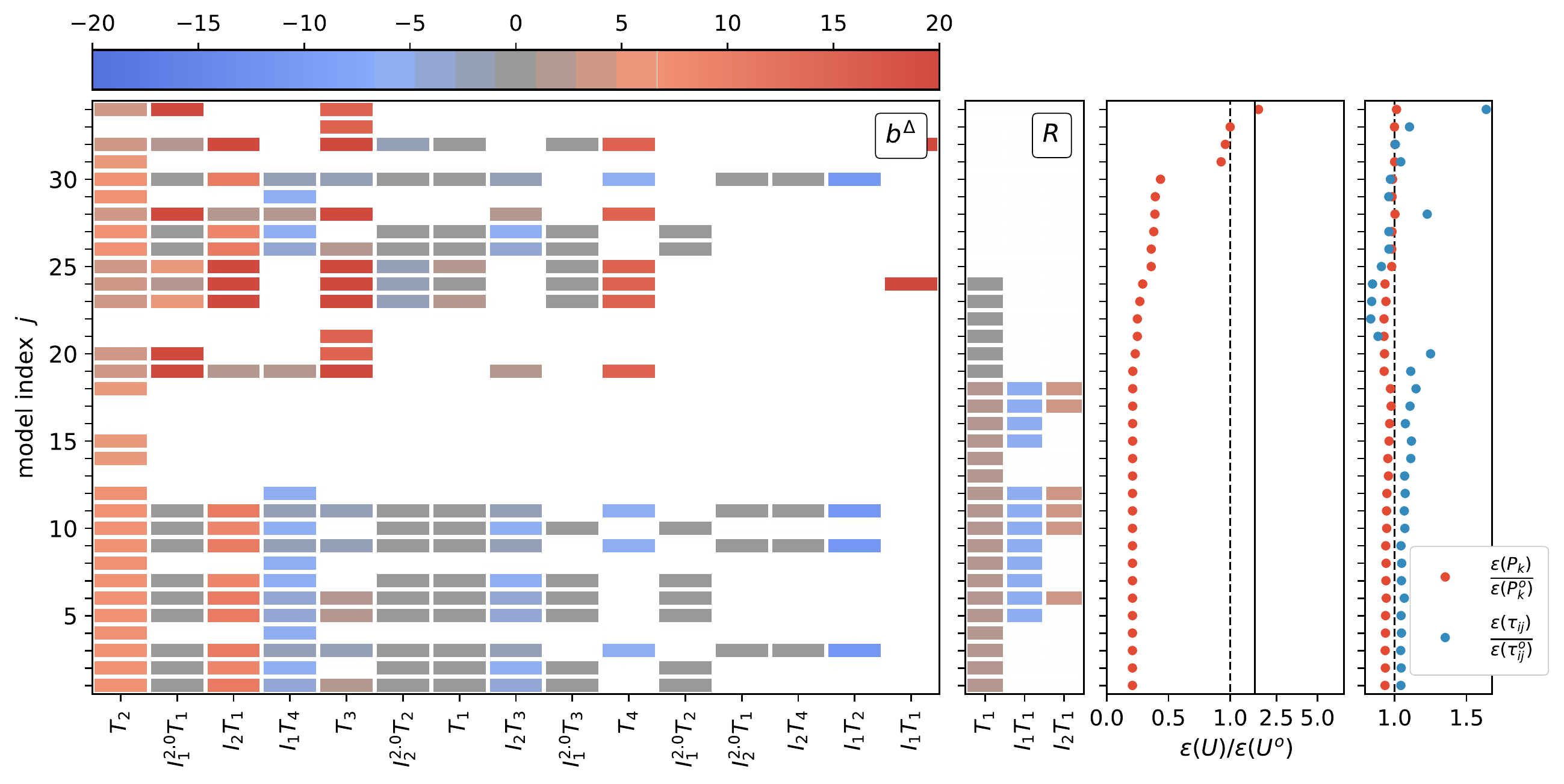}
    }
	\hfill
    \subfloat[\cbfs{} \label{fig:modelstruc:cbfs}]{%
		\includegraphics[width=\textwidth]{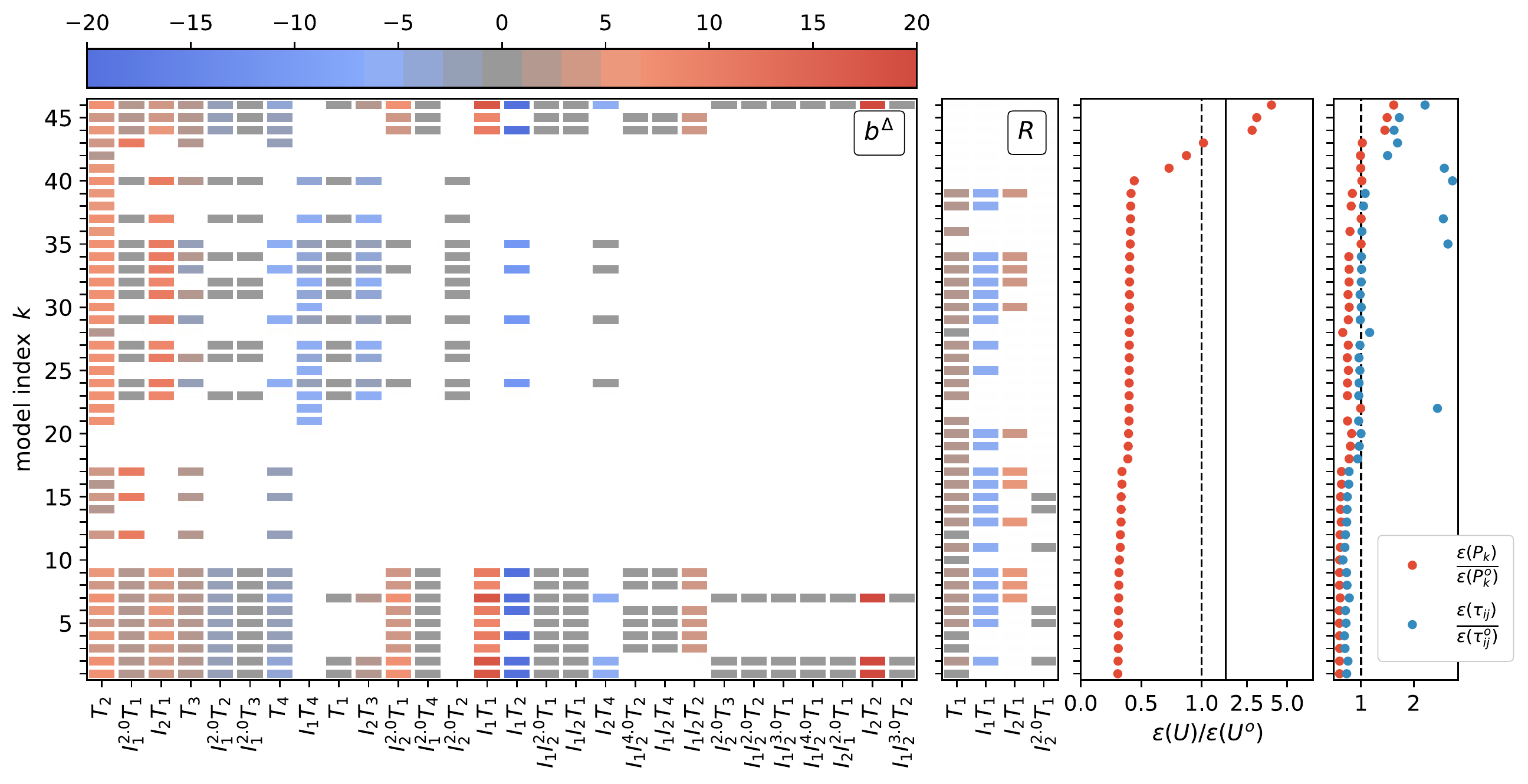}
    }
    \caption{The two matrices (l.) show the models $M_i$ for $\bdelta$ and $R$. The mean-squared error in velocity $U$, production $P_k$ and Reynolds-stress $\tau_{ij}$ normalised by the mean-squared error of the baseline \komegasst{} model is also shown (mid to right).}
    \label{fig:modelstruc:post}
\end{figure}
\subsection{Cross-validation using CFD}
\begin{table}
\caption{Best-predictive models with rank (index $i,j,k$ in Figure \ref{fig:modelstruc:post}) and normalised error on velocity $\epsilon(U)/\epsilon(U^o)$ for different cases.}
\label{tab:models}       
\begin{tabular}{lllllll}
\hline\noalign{\smallskip}
 & \multicolumn{2}{c}{\ph{}} & \multicolumn{2}{c}{\cd{}} & \multicolumn{2}{c}{\cbfs{}} \\
Model & index $i$ & $\epsilon(U)/\epsilon(U^o)$ & index $j$ & $\epsilon(U)/\epsilon(U^o)$ & index $k$ & $\epsilon(U)/\epsilon(U^o)$ \\
\noalign{\smallskip}\hline\noalign{\smallskip}
$M^{(1)}$ & (1.) & 0.22287 & (19.) & 0.21146 & - & 0.30413 \\
$M^{(2)}$ & - & 0.38867 & (1.) & 0.20828 & (26.) & 0.40154  \\
$M^{(3)}$ & (3.) & 0.22744 & - & 0.22422  & (1.) & 0.30655  \\
\noalign{\smallskip}\hline
\end{tabular}
\end{table}
Cross-validation tests how well models identified on training data perform on unseen test cases \cite{Bishop2006}. This assessment allows to determine the best-predictive models from a set. As stated above, the role of the frozen, training data should not be overcharged, so that we cross-validate using CFD. By doing so, we can assess the validity of \method{} as a tool for model discovery as well as the predictive performance of the identified models outside of their training set.

The selected correction models regress $\bdelta$ and $R$ individually and can also be applied individually for predictions when implemented in the solver, i.e. a model correcting $\bdelta$ can be used without a correction of $R$ and vice-versa. This gives us 8 models per training data for \ph{} and \cd{} and 6 for \cbfs. Also, we can study their combined effect. With $5$ models for $\bdelta$ and $3$ for $R$ we have additional 15 possible combinations, which makes in total 23 distinct models for the training data of \ph{} and \cd{} and 11 distinct model combinations for the training data of \cbfs. For the cross-validation in the following, we conduct in total 35 for test cases \ph{} and \cd{} and 47 simulations for test case \cbfs{} including the baseline simulation with the uncorrected \komegasst{}. 

In Figure \ref{fig:U:MSE} the mean-squared error of each model on the velocity field $\epsilon(U)$ normalised with the mean-squared error of the baseline $\epsilon(U^o)$ is shown. The type of model, whether it is providing a correction both for $\bdelta$ and $R$ or for each one individually, and from which training data it originated from, is emphasised by a unique marker form and color combination. Whether  the correction for $\bdelta$ needs to be scaled with $\xi=0.1$ to achieve convergence, see Section \ref{sec:inference}, is indicated by a black marker edge. Most of the models show a good or even substantial improvement over the baseline. But, for the set of models, only providing a correction for $\bdelta$, most but not all lead to an improvement of the resulting velocity field. In contrast to that, if only a correction for $R$ is deployed, the result is a consistent, substantial improvement across all test cases. Using both a model for $\bdelta$ and $R$ leads to a further improvement, except for test case \cbfs{}. Surprisingly, the best model per test case is not always identified on the associated training data. While this expectation holds for the cases \cbfs{} and \cd{} it is not true for \ph{}, for which the other two training sets deliver significantly better performing models. In general, the data of \cd{} and \cbfs{} provide models, which are well performing on all test cases presented.

In Figure \ref{fig:modelstruc:post}, both the error and the model structure for the correction of $\bdelta$ as well as for $R$ is shown. The models are ordered according to the mean-squared error on the stream-wise velocity $U$. In line of the discussion of Figure \ref{fig:U:MSE} three groups can be identified: a few models, which lead to an increased error compared to the baseline; a small group of models per test case, which are equal or similar to the baseline; and the great majority of models, which result in an improvement. It can be observed how the error in the velocity is significantly reduced once a correction of $R$ is used. The two other error plots in Figure \ref{fig:modelstruc:post} give an indication of the relative performance of the models compared to the baseline. The first is the mean-squared error of the total production $P_k$ within the $k$ equation and the second one is the mean-squared error of the Reynolds-stress $\tauij$ normalised by the baseline result. Following the rationale of correcting terms with the baseline model, improving these terms should lead to an improved velocity. For the cases \cd{} and \cbfs, we see that the error for $U$ and $P_k$ reduce simultaneously for most of the models. Also the error in $\tauij$ shows a reduction when the error in $U$ decreases, but not as significant as the error in $P_k$. For a group of models between model index $5<j<20$ for case \cd{}, we see a jump in the error in $P_k$ and $\tauij$, while the error in the velocity is not changing compared to the neighbouring models. For case \ph{}, we also observe a strong reduction of the $P_k$ error for an active $R$ correction. Surprisingly, for these the error in $\tauij$ increases. When the error in $U$ is further reduced the error in $\tauij$ also decreases, but the error in $P_k$ increases again. It can be observered, that the best models correct the velocity up to 5 times better in mean-squared error than the \komegasst{} baseline model. This leaves still room for further improvement compared to the error using the frozen data sets, see Table \ref{tab:verify:mse}. But, especially for case \cbfs{} the result is already very close to the possible correction provided by the frozen data at least for $U$.

\begin{figure}
	\centering
	\subfloat[\ph{} \label{fig:U:ph}]{%
       \includegraphics[width=0.85\textwidth]{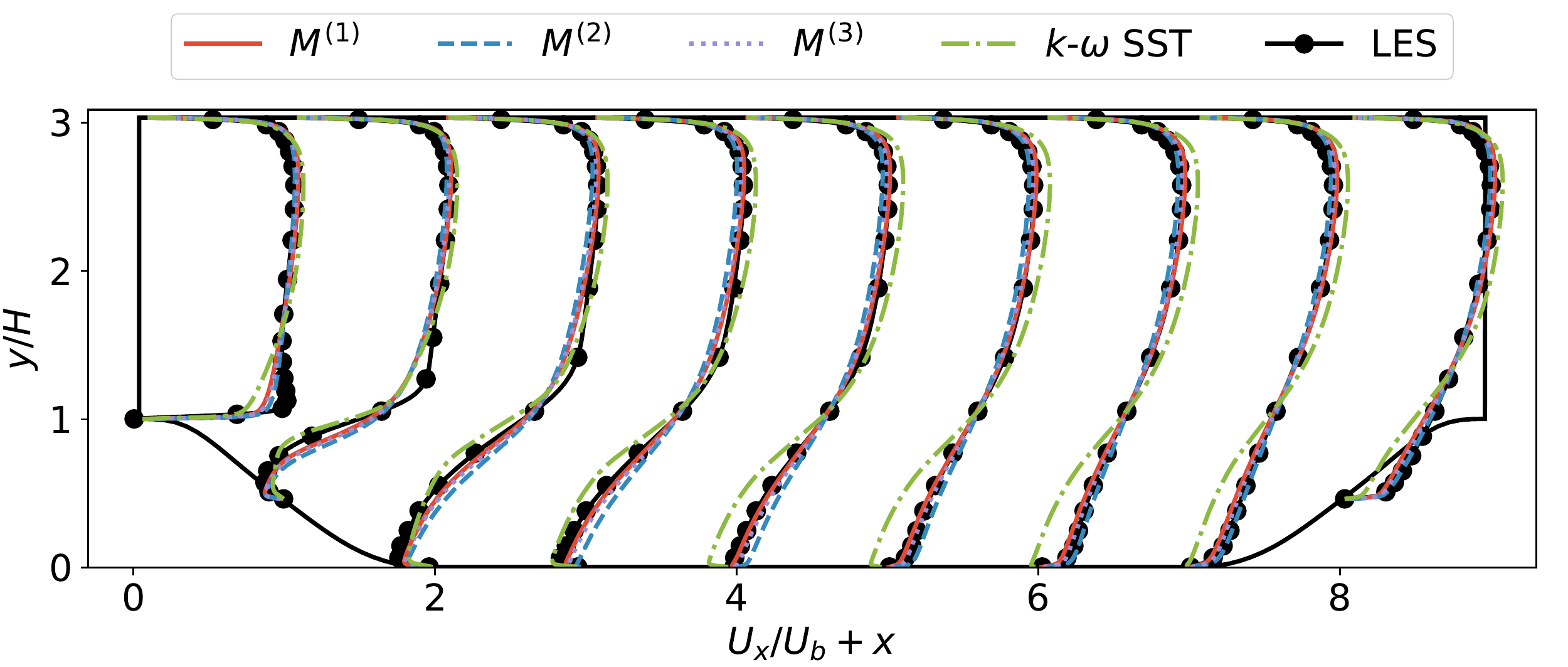}
    }
    \hfill
    \subfloat[\cd{} \label{fig:U:cd}]{%
		\includegraphics[width=0.85\textwidth]{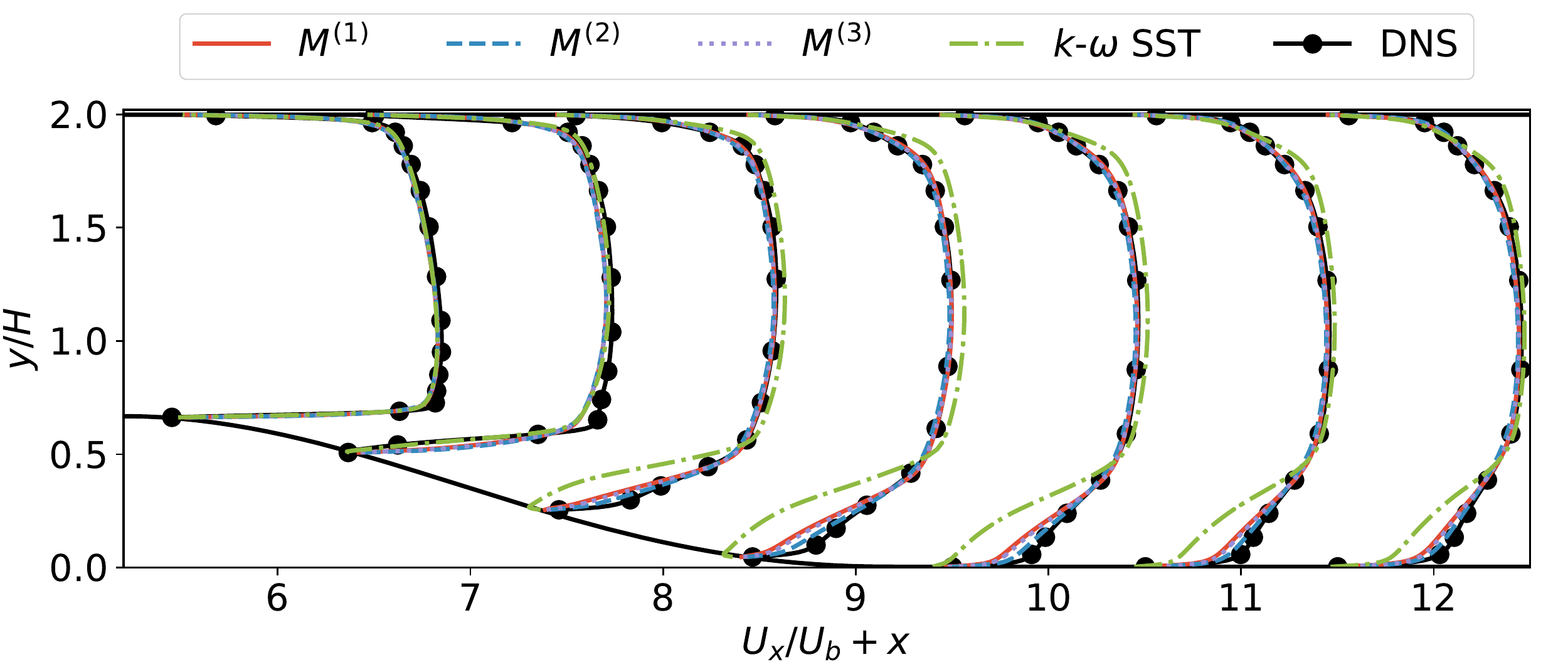}
    }
	\hfill
    \subfloat[\cbfs{} \label{fig:U:cbfs}]{%
		\includegraphics[width=0.85\textwidth]{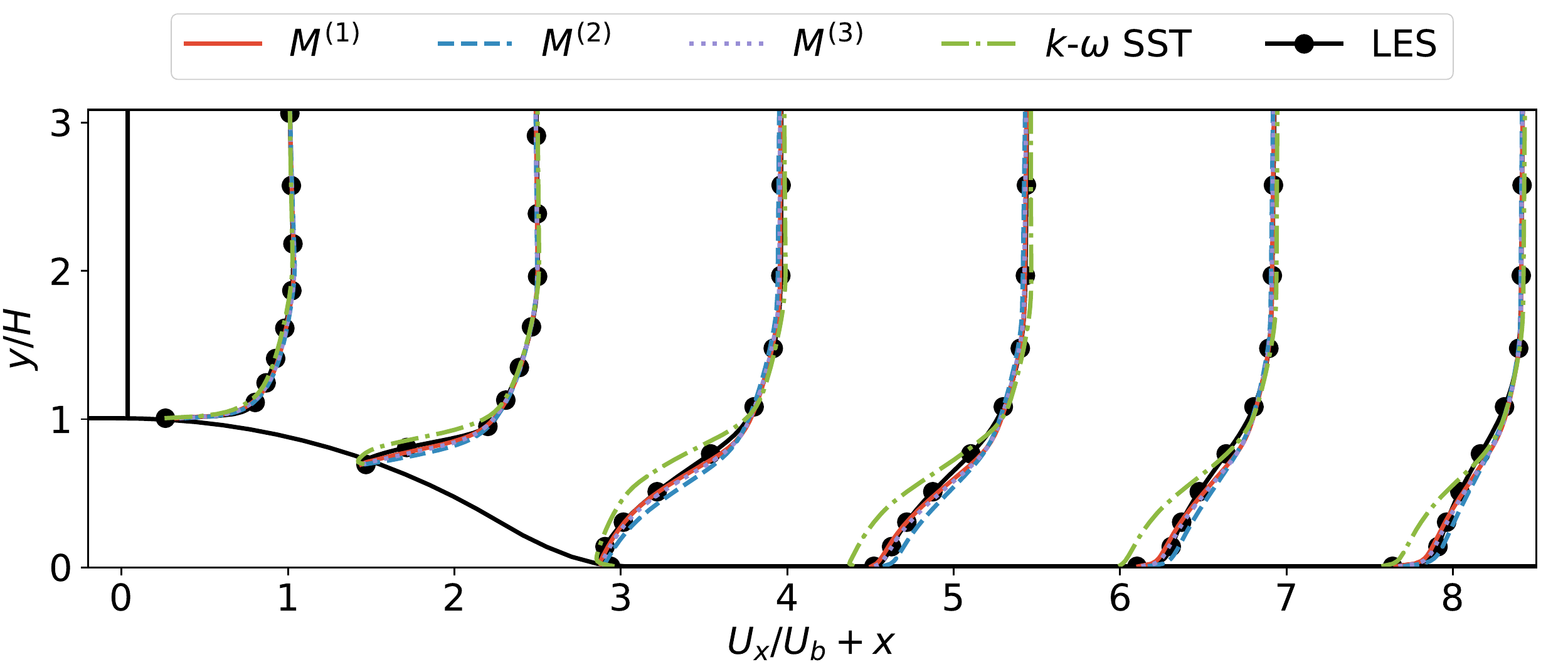}
    }
    \caption{Predicted stream-wise velocity.}
    \label{fig:U}
\end{figure}

\begin{figure}
	\centering
	\subfloat[\ph{} \label{fig:k:ph}]{%
       \includegraphics[width=0.85\textwidth]{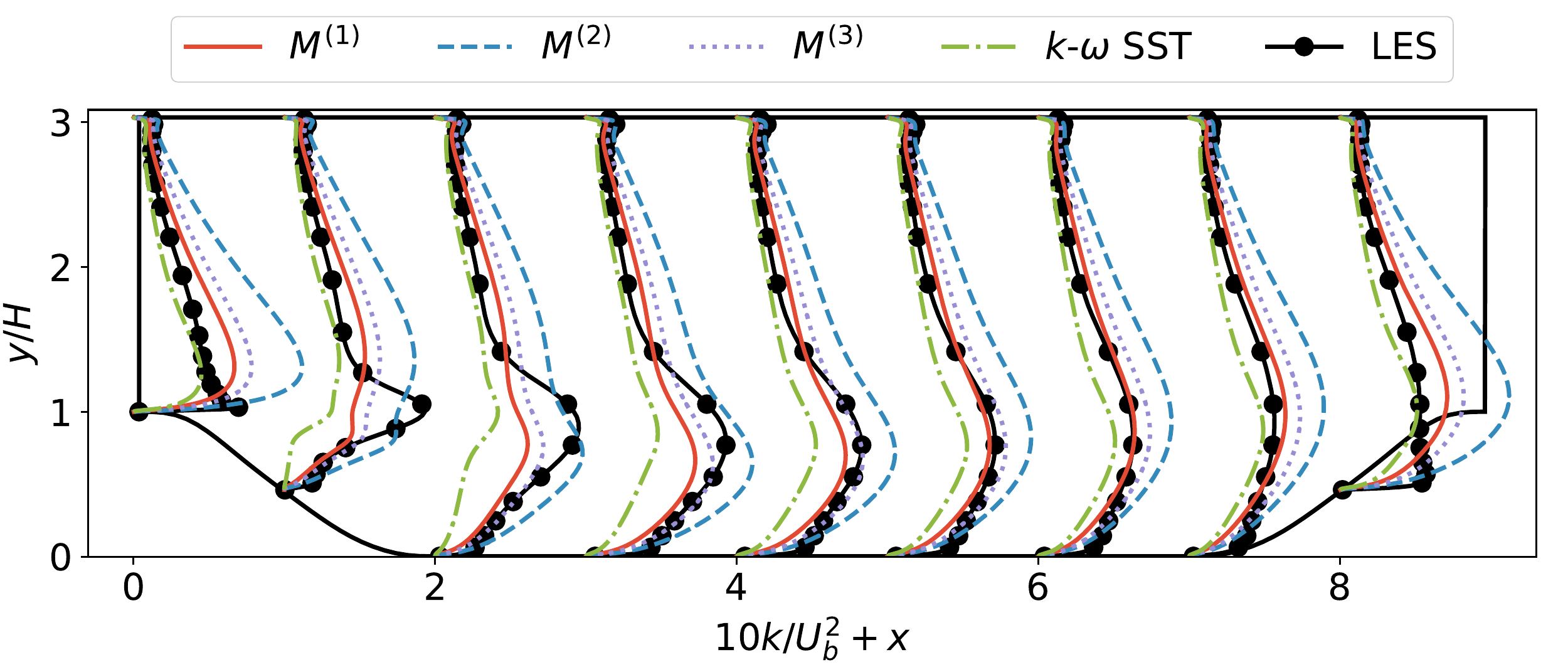}
    }
    \hfill
    \subfloat[\cd{} \label{fig:k:cd}]{%
		\includegraphics[width=0.85\textwidth]{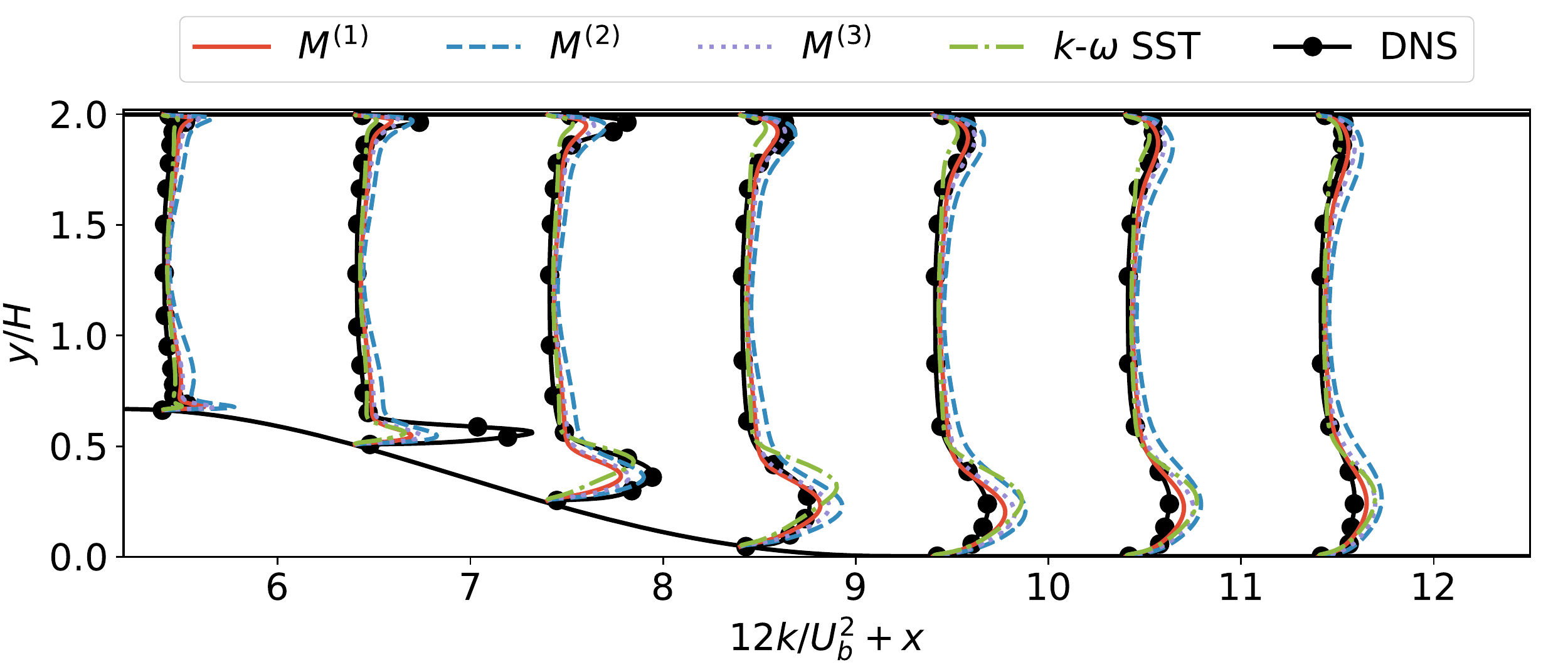}
    }
	\hfill
    \subfloat[\cbfs{} \label{fig:k:cbfs}]{%
		\includegraphics[width=0.85\textwidth]{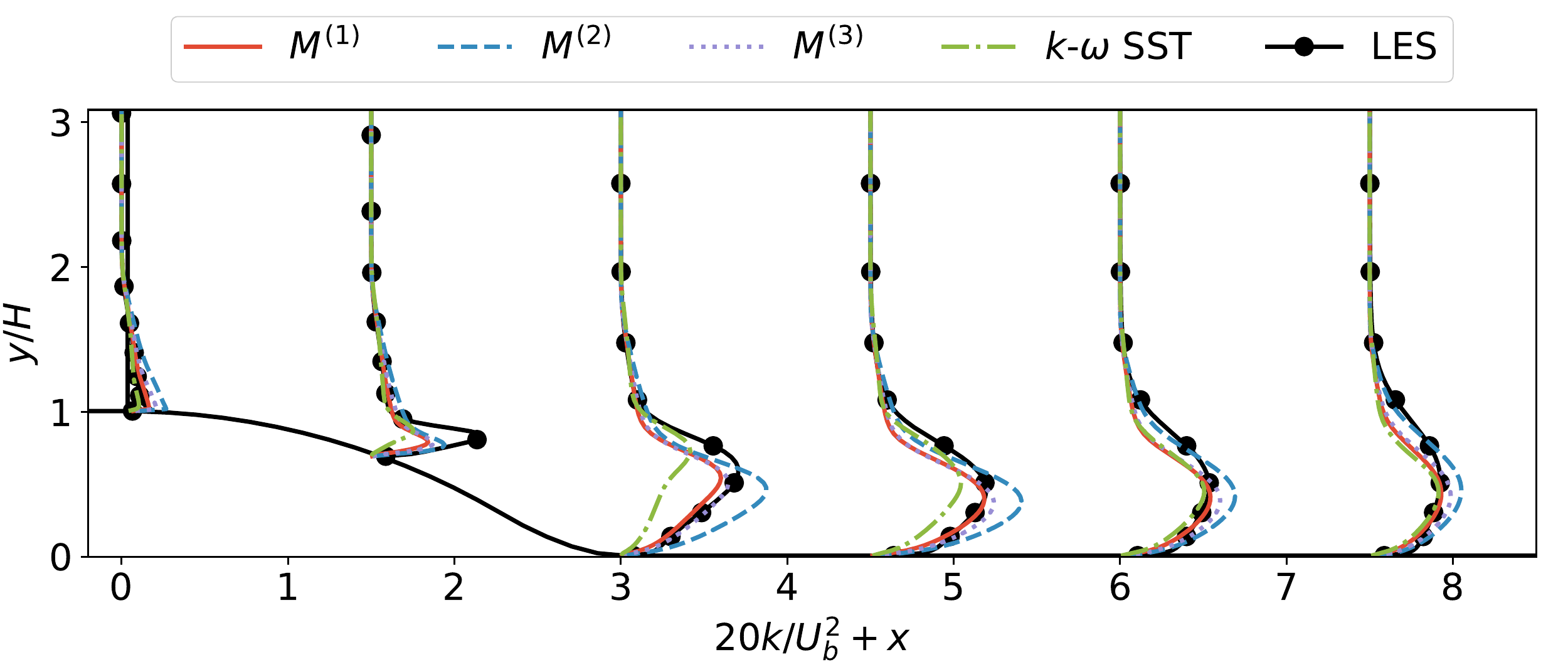}
    }
    \caption{Predicted turbulent kinetic energy.}
    \label{fig:k}
\end{figure}

\begin{figure}
	\centering
	\subfloat[\ph{} \label{fig:tauxy:ph}]{%
       \includegraphics[width=0.85\textwidth]{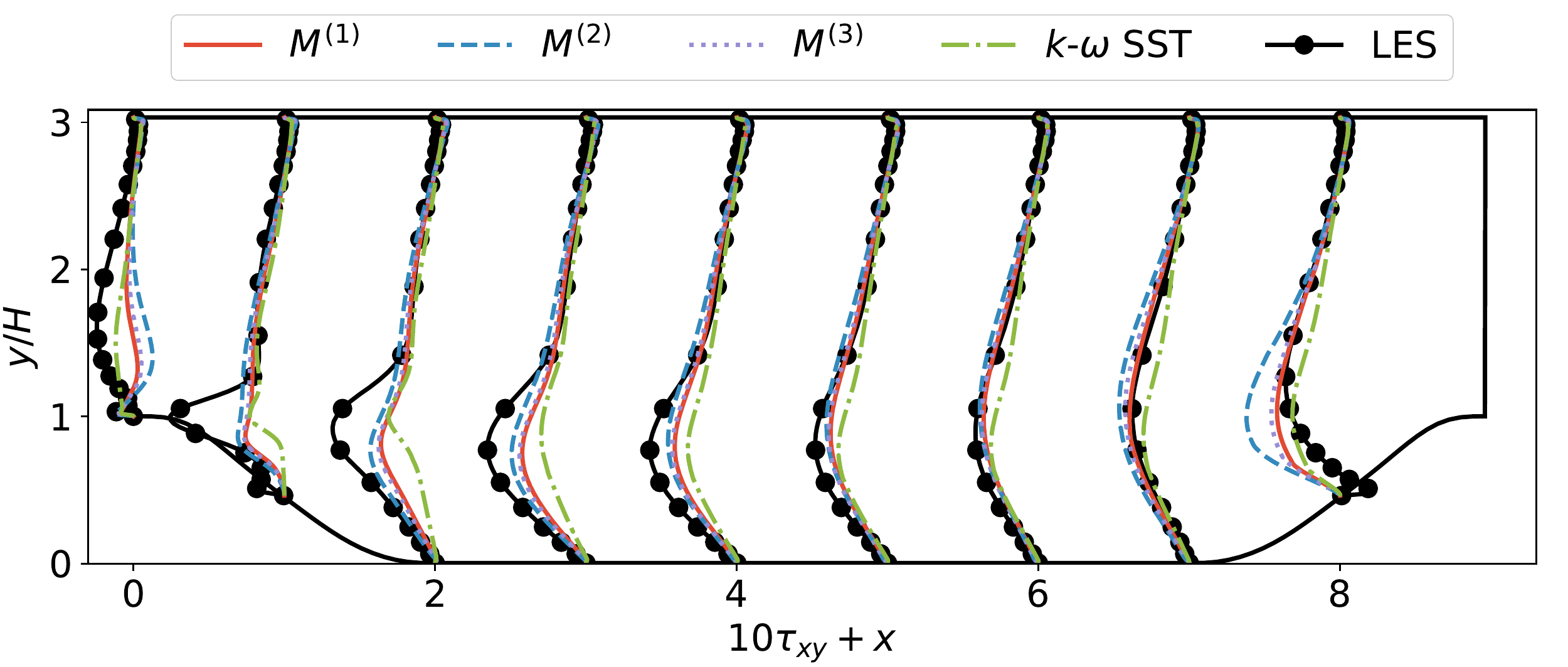}
    }
    \hfill
    \subfloat[\cd{} \label{fig:tauxy:cd}]{%
		\includegraphics[width=0.85\textwidth]{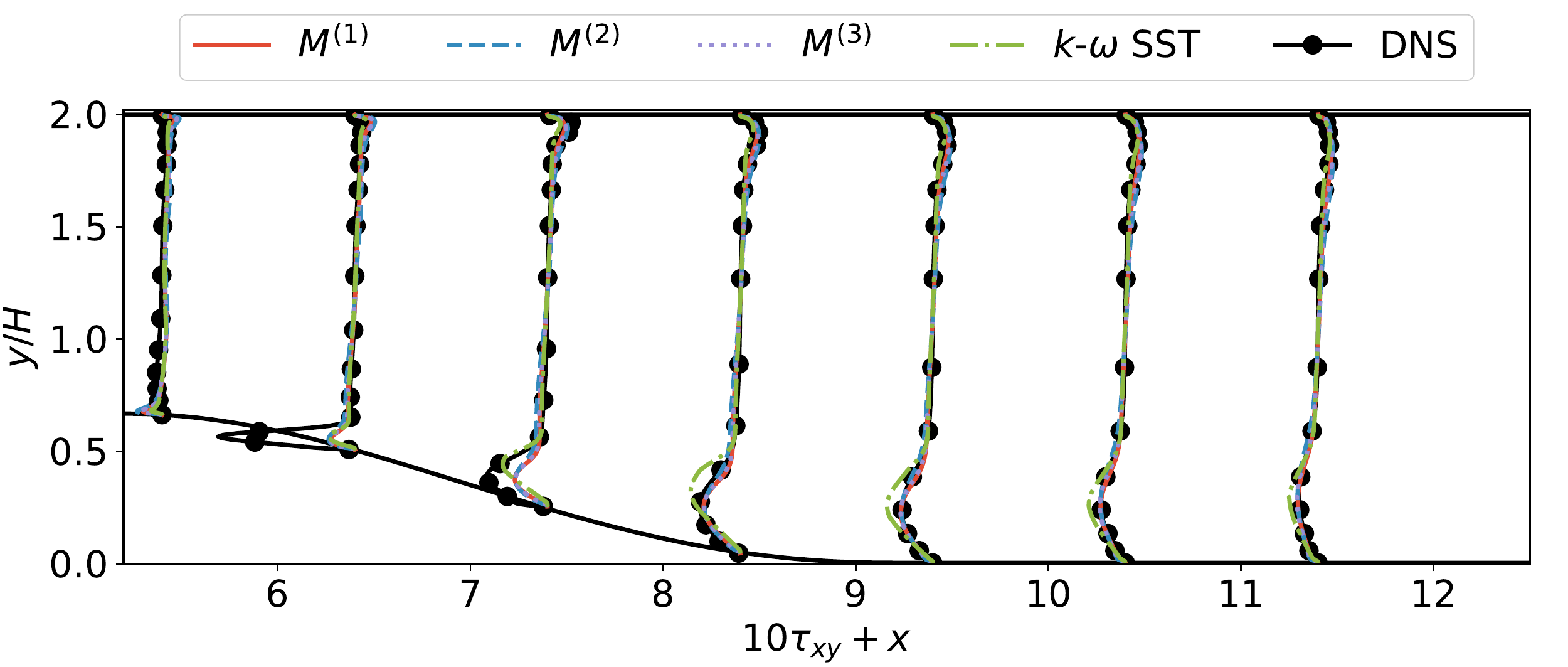}
    }
	\hfill
    \subfloat[\cbfs{} \label{fig:tauxy:cbfs}]{%
		\includegraphics[width=0.85\textwidth]{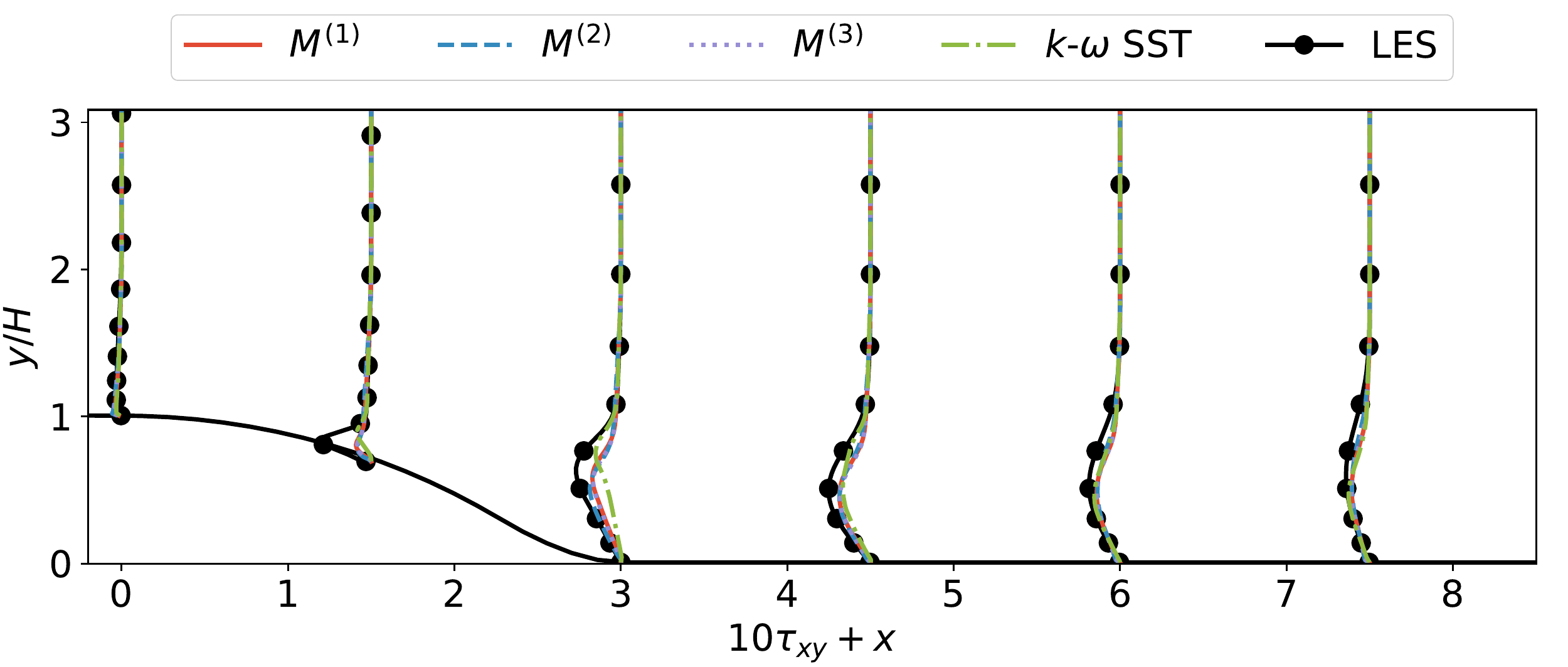}
    }
    \caption{Predicted shear stress.}
    \label{fig:tauxy}
\end{figure}

\begin{figure}
	\centering
	\subfloat[\ph{} \label{fig:wss:ph}]{%
       \includegraphics[width=0.85\textwidth]{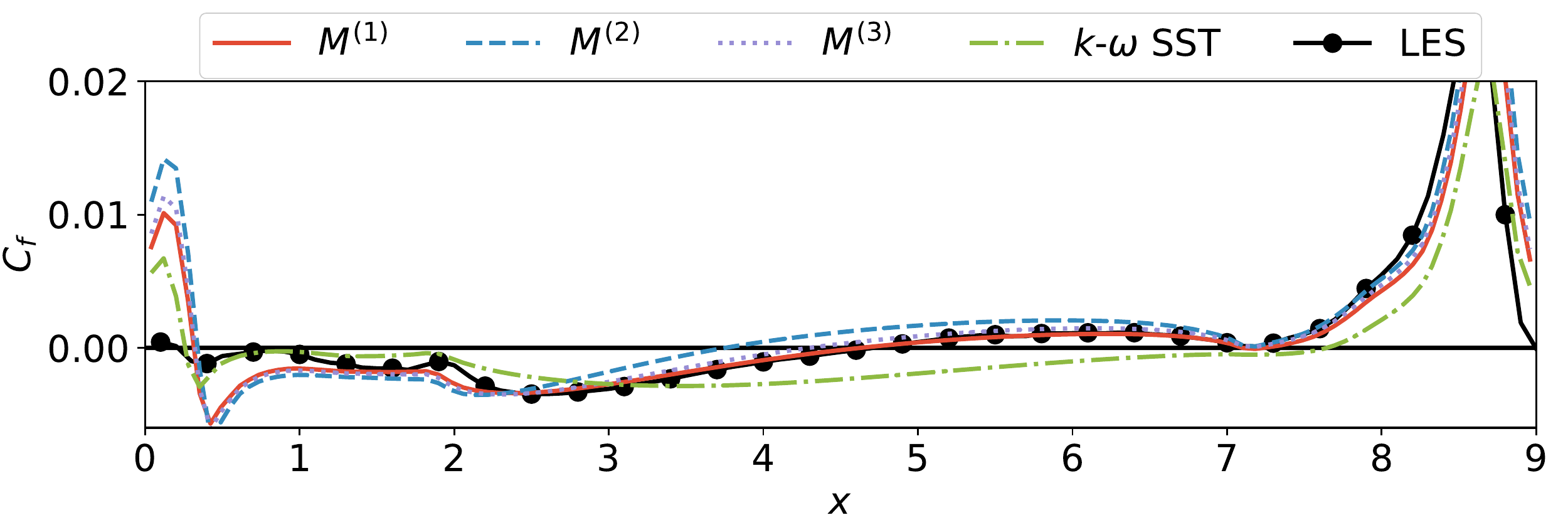}
    }
    \hfill
    \subfloat[\cd{} \label{fig:wss:cd}]{%
		\includegraphics[width=0.85\textwidth]{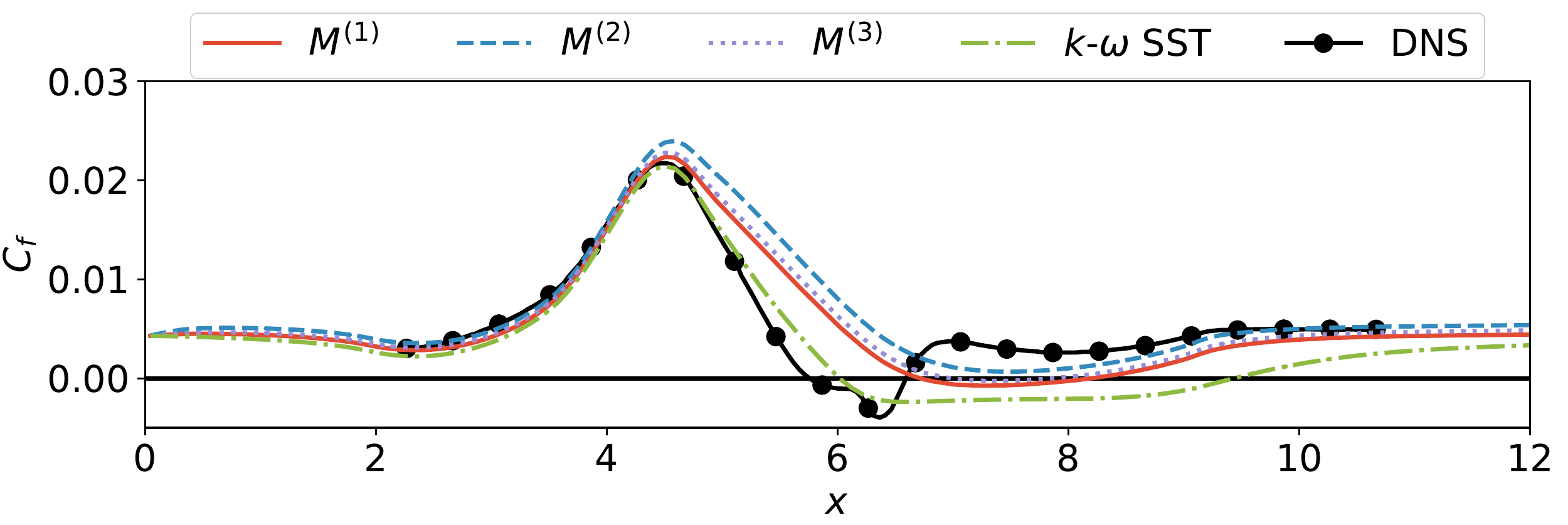}
    }
	\hfill
    \subfloat[\cbfs{} \label{fig:wss:cbfs}]{%
		\includegraphics[width=0.85\textwidth]{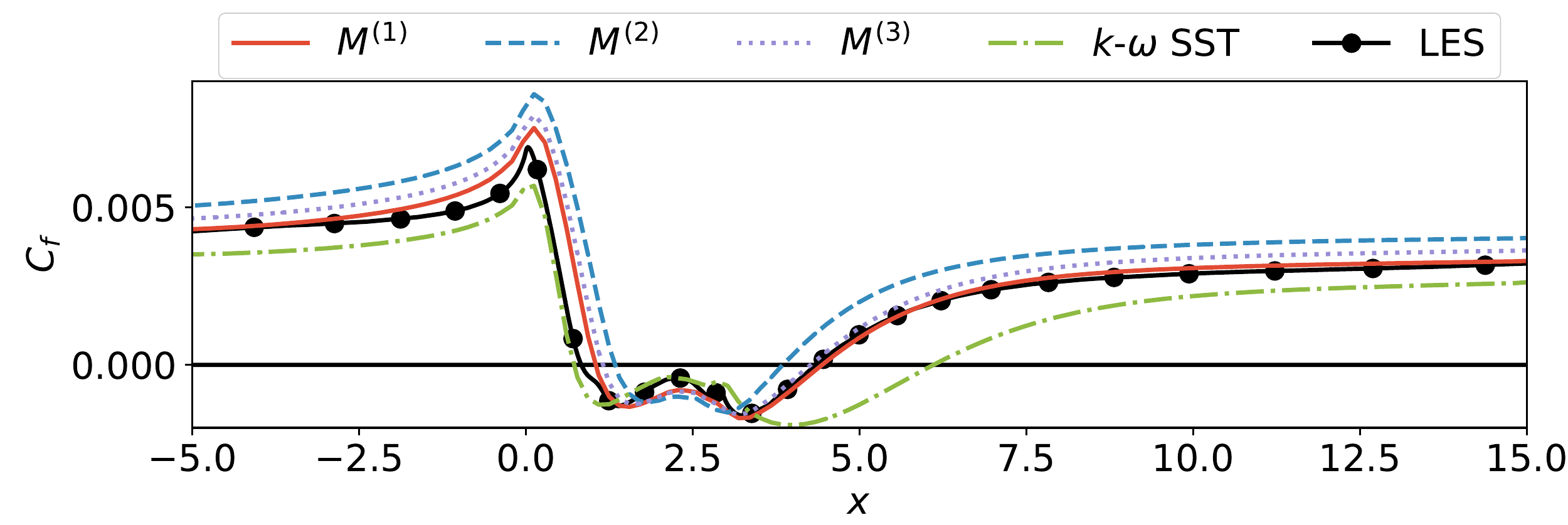}
    }
    \caption{Predicted skin friction coefficient.}
    \label{fig:wss}
\end{figure}

Given this cross-validation assessment we select models $\bm{M}^{(i)} = (M^{(i)}_{b^\Delta}, M^{(i)}_{R} )^T$ based on the lowest $\epsilon(U)$ per case
\begin{align}
	\m{(1)}{b^\Delta} = & \; \left(24.94 I_{1}^{2} + 2.65 I_{2}^{1}\right) \T{1} + 2.96 \; \T{2}, \nonumber \\
	&+ \left(2.49 I_{2}^{1} + 20.05\right) \T{3} + \left(2.49 I_{1}^{1} + 14.93\right) \T{4},  \nonumber\\	
 	\m{(1)}{R} = & \; 0.4 \; \T{1}, 
 \end{align}
 
 \begin{align}
 	\m{(2)}{b^\Delta} = & \; \T{1} \left(0.46 I_{1}^{2} + 11.68 I_{2}^{1} - 0.30 I_{2}^{2} + 0.37\right)  \nonumber\\
 	&+ \T{2} \left(- 12.25 I_{1}^{1} - 0.63 I_{2}^{2} + 8.23\right) \nonumber\\
 	&+ \T{3} \left(- 1.36 I_{2}^{1} - 2.44\right) \nonumber\\
	&+ \T{4} \left(- 1.36 I_{1}^{1} + 0.41 I_{2}^{1} - 6.52\right), \nonumber\\
	\m{(2)}{R} = &\; 1.4 \; \T{1},
	\label{eq:M2}
\end{align}

\begin{align}
 	\m{(3)}{b^\Delta} = & \; T_{1} \Bigl( 0.11 I_{1}^{1} I_{2}^{1} + 0.27 I_{1}^{1} I_{2}^{2} - 0.13 I_{1}^{1} I_{2}^{3} + 0.07 I_{1}^{1} I_{2}^{4}\nonumber \\ 
 	& + 17.48 I_{1}^{1} + 0.01 I_{1}^{2} I_{2}^{1} + 1.251 I_{1}^{2} + 3.67 I_{2}^{1} + 7.52 I_{2}^{2} - 0.3 \Bigr) \nonumber\\
 	&+ T_{2} \left(0.17 I_{1}^{1} I_{2}^{2} - 0.16 I_{1}^{1} I_{2}^{3} - 36.25 I_{1}^{1} - 2.39 I_{1}^{2} + 19.22 I_{2}^{1} + 7.04\right) \nonumber\\
&+ T_{3} \left(- 0.22 I_{1}^{2} + 1.8 I_{2}^{1} + 0.07 I_{2}^{2} + 2.65\right) \nonumber\\
&+ T_{4} \left(0.2 I_{1}^{2} - 5.23 I_{2}^{1} - 2.93\right), \nonumber\\
	\m{(3)}{R} = &\; 0.93 \; \T{1},
\end{align}
\noindent for which further details on the corresponding training data and the rank of the model on each test case are given in Table \ref{tab:models}. Especially model $\m{(3)}{}$ performs very well both on \cbfs{} (rank 1.) and \ph{} (3.). While the rank of the others varies more between the test cases, they are still within the set of well-performing models with $\epsilon(U)/\epsilon(U^o)<0.5$. Their predictions of stream-wise velocity $U$, $k$, the Reynolds-stress component $\tau_{xy}$ and the skin-friction coefficient $C_f$ are shown in Figure \ref{fig:U} to \ref{fig:wss} for the three test cases. As already stated for the error evaluated on the entire domain discussed above, these three models show an improvement of the spatial distribution of the predicted quantities in comparison to the baseline prediction of \komegasst. Especially the velocity is well-captured for all three. While $k$ is better identified compared to the baseline, we still observe a discrepancy between the predictions and the data. For \ph{} the three models do not fit the complex spatial structure especially in the shear-layer, but together encapsulate the data for most of the profiles. For \cd{} the models are underestimating $k$ for $x<7$ and overestimate it further downstream. For \cbfs{} the models also underestimate on the curved surface, but fit the data better than the baseline for $3<x<5$. The magnitude of the Reynolds-stress component $\tau_{xy}$ is underestimated on the curved surfaces of all test cases. For \ph{} the models fail to fit the complex spatial structure especially within the separated shear-layer behind the hill and on the hill itself. The skin friction coefficient $C_f$ and the associated separation and reattachment points are best captured by $\m{(1)}{}$ and $\m{(3)}{}$ for \ph{} and \cbfs{} and systematically under-estimated with $\m{(2)}{}$, i.e.~a shorter recirculation zone. For \cd{}, we observe a small recirculation zone as reported in the literature, but too far down-stream.  However, the baseline \komegasst{} drastically over-predicts this zone and the model $\m{(2)}{}$ ignores it entirely.

Overall, the models $\m{(1)}{}$ and $\m{(3)}{}$ agree best with the data, which is in line with the global error on $U$ in Table \ref{tab:models}. The models are different in their form, but show similar error values and spatial structure across the test cases. Model $\m{(2)}{}$ tends to overestimate the magnitude of the quantities $U$, $k$ and $\tau_{xy}$ and therefore predicts smaller or no separation bubbles. This model was identified using \ph{} as training data and, ignoring the specific structure for $\bdelta$, has the largest coefficient for correcting $R$, see \eqref{eq:M2}, which leads to larger $k$ compared to the others, which is the reason for the systematic over-prediction. 

\begin{figure}
\centering
  \includegraphics[width=0.85\textwidth]{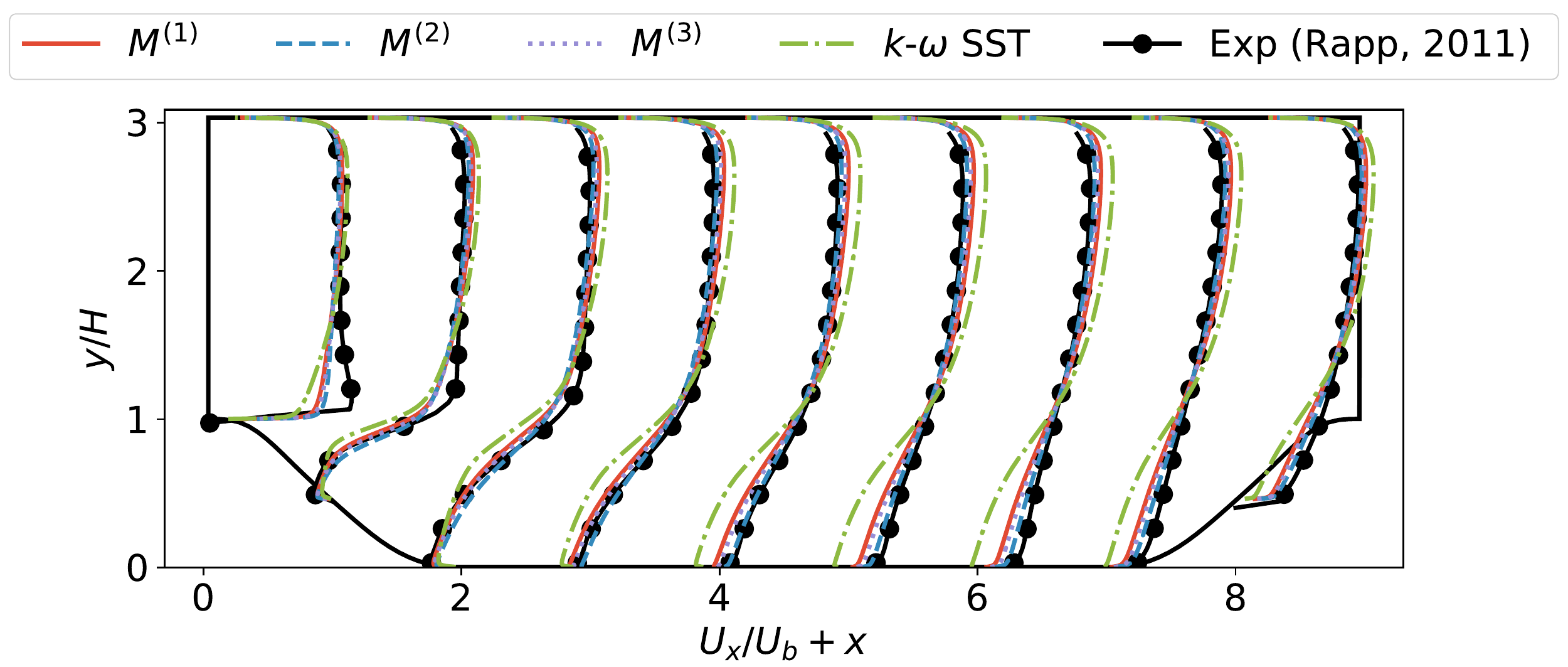}
\caption{Flow over periodic hills at $Re=37000$ using correction models compared to baseline \komegasst{} and experimental data of \cite{Rapp2011}.}
\label{fig:U:ph37000}       
\end{figure}

In order to test how the models extrapolate to cases of larger $Re$, we predict the flow over periodic hills at $Re=37000$, see Figure \ref{fig:U:ph37000}. Due to an increase of turbulence this case has a significantly shorter recirculation zone. For this \emph{true} prediction throughout the domain the three models improve significantly compared to the baseline. Interestingly, the model $\m{(2)}{}$ is delivering the best fit of the data and the others tend to slightly underestimate it. Thus, taking the results of the cross-validation on the low-$Re$ cases into account, the models show a weak $Re$-dependence, but overall robustness between the cases.

\section{Conclusion and extension}
In this work \method{} was introduced to discover algebraic models in order to correct the model-form error within the \komegasst{}. For this novel machine learning method two additive terms, on the level of the stress-strain relation $\bdelta$ and within the turbulent transport equations $R$, were identified by means of \frozen{}, for which the governing equations are evaluated given high-fidelity data of three cases of separating flows. It was validated that the computed terms are compensating the model-form error and reproduce the high-fidelity LES or DNS mean-flow data. Hence, \frozen{} is a cost-efficient way to distill useful information directly from full-field data without the need of an inversion procedure.

Cross-validation of the discovered models using CFD was carried out to rank the models. While using both corrections for $R$ as well es for $\bdelta$ lead to a systematic improvement of the predictions over the baseline, a correction only for $R$ can already be enough to achieve sufficient results for the velocity field. For the best performing models on each case both the global error on $U$ as well as the spatial structure on $U$, $k$ and $\tau_{xy}$ was coherent. The models also performed well for the periodic hills flow at a much larger $Re$-number ($Re=37000$). As the sparse regression is computationally inexpensive, \method{} allows for rapid discovery of robust models, i.e.~a model trained for one flow may perform well for flows outside of the training range, but with similar features.

Overall, the present systematic study has shown the capabilities of \method{} to discover effective corrections to \komegasst{}. Further work will focus on making the model-filtering and the inference step of \method{} more systematic and data-driven. We will also apply \method{} to a larger variety of flow cases in order to show its potential for rapid model discovery of corrections for industrial purposes.

\section*{acknowledgements}
The authors wish to thank Richard Sandberg for sharing OpenFOAM code for comparison of implementations and Michael Breuer for providing the full-field LES and DNS simulation data for the periodic hill flow case.

%
%
\section*{Compliance with Ethical Standards}
\small{\noindent \textbf{Funding} \hspace{5pt} This research has received funding from the European Unions Seventh Framework Programme under grant number ACP3-GA-2013-605036, UMRIDA project.

\noindent \textbf{Conflict of Interest} \hspace{5pt} The authors declare that they have no conflict of interest.
}
%
\bibliographystyle{spphys}       
%
%
%

\end{document}